\documentclass[journal,twocolumn]{IEEEtran}

\usepackage{algorithm,algpseudocode,amsbsy,epsfig,bbm,mathrsfs,multirow}
\usepackage[T1]{fontenc}

\usepackage{amsmath,amsxtra,amssymb,latexsym, amscd,amsthm}

\usepackage[latin9]{inputenc}
\usepackage{amsmath}
\usepackage{fixltx2e}

\usepackage{array,multirow,graphicx}
\usepackage{color}
\usepackage{cite}
\usepackage{float}
\usepackage{makecell}
\usepackage[x11names,dvipsnames,table]{xcolor} 
\usepackage{colortbl}
\usepackage{multirow}
\usepackage{graphicx}

\usepackage{tabularx}

\usepackage{caption}
\usepackage{subcaption}

\newcommand{\LineComment}[1]{\Statex \hfill\textit{#1}}

\usepackage{lipsum}
\definecolor{mygray}{gray}{0.6}
\definecolor{myblue}{rgb}{0.8,0.85,1} 
\usepackage{array}
\newcolumntype{L}[1]{>{\raggedright\let\newline\\\arraybackslash\hspace{0pt}}m{#1}}
\newcolumntype{C}[1]{>{\centering\let\newline\\\arraybackslash\hspace{0pt}}m{#1}}
\newcolumntype{R}[1]{>{\raggedleft\let\newline\\\arraybackslash\hspace{0pt}}m{#1}}

\usepackage{array, tabularx, boldline}
\usepackage{eurosym}
\usepackage{amstext} 
\DeclareRobustCommand{\officialeuro}{%
  \ifmmode\expandafter\text\fi
  {\fontencoding{U}\fontfamily{eurosym}\selectfont e}}

\usepackage{array, tabularx, boldline}
\usepackage{graphicx}
\usepackage{cellspace}
\newcommand{\bx}{\ensuremath{{\bf x}}}

\newcommand{\bu}{\ensuremath{{\bf u}}}

\newcommand{\opt}{\ensuremath{\mathsf{OPT}}}

\newcommand{\fptas}{\ensuremath{\mathsf{FPTAS}}}
\newcommand{\greedy}{\ensuremath{\mathsf{GREEDY}}}

\newcommand{\sqp}{\ensuremath{\mathsf{SQP}}}
\newcommand{\pof}{\ensuremath{\mathsf{POF}}}
\newcommand{\poe}{\ensuremath{\mathsf{POE}}}
\newcommand{\tim}{\ensuremath{\mathsf{TIME}}}

\newtheorem{theorem}{Theorem}

\newtheorem{lemma}{Lemma}

\newtheorem{claim}{Claim}

\def\cF{\mathcal F}

\def\cI{\mathcal I}

\def\cN{\mathcal N}
\def\cS{\mathcal S}

\renewcommand{\algorithmicrequire}{\textbf{Input:}}
\renewcommand{\algorithmicensure}{\textbf{Output:}}

\begin{document}
\title{Access Management in Joint Sensing and Communication Systems: Efficiency versus Fairness}
\author{Trung Thanh Nguyen, Khaled Elbassioni, Nguyen Cong Luong, Dusit Niyato,~\IEEEmembership{Fellow,~IEEE}
and
Dong In Kim,~\IEEEmembership{Fellow,~IEEE}
}

\maketitle
\begin{abstract}
In this paper, we consider a distributed joint sensing and communication (DJSC) system in which multiple radar sensors are deployed. Each radar sensor is equipped with a sensing function and a communication function, and thus it is considered to be a JSC node. The JSC nodes are able to perform sensing their surrounding environments, e.g., weather conditions or available spectrum. Furthermore, they can cooperatively detect and track a common target. The information, i.e., of the environment and target, collected by the JSC nodes is transmitted to a base station (BS), i.e., a data fusion point, for further processing. As such, different aspects of the target to be viewed simultaneously, which significantly improves the performance of the target detection and tracking. 
However, both the sensing function and communication function require a certain amount of bandwidth for their operations, and deploying multiple JSC nodes may consume a large amount of bandwidth. Therefore, we investigate the bandwidth allocation problem for the DJSC system. In particular, we aim to optimize the bandwidth allocation to the sensing function and the communication function of the JSC nodes. To improve the allocation efficiency while benefiting the spatial diversity advantage of the DJSC systems, the objective is to maximize the sum of sensing performances, i.e., estimation rates, communication performances, i.e., communication data rates, and fairnesses of all the users. The optimization problem is non-convex and difficult to be solved. For this, we propose a fully polynomial time approximation algorithm, and we prove that the approximation algorithm can guarantee a near-optimal solution with an accuracy bound of $\epsilon$. Furthermore, we propose to use a heuristic algorithm with lower complexity. The simulation results show that both the proposed algorithms are able to achieve the solutions close to the optimum in a computationally efficient fashion. 

\end{abstract}
\section{Introduction}
\label{sec:intro}

Joint sensing and communication (JSC) has been of significant interest in recent years due to its important benefits. Firstly, JSC enables sensing and communication systems to share spectrum bands with each other, and thus significantly improving the spectrum utilization. Secondly, JSC allows a single hardware platform (such as a UAV or an
autonomous vehicle) to concurrently execute both the sensing function and the
communication function. As such, JSC improves the efficiency of resources, i.e., spectrum and energy, reduces the system size, and minimizes the system cost. These advantages make JSC become one of the most potential technologies for civilian applications, e.g., autonomous vehicle systems and flying wireless mesh networks~\cite{ma2020joint}, and military applications, e.g., flying target tracking~\cite{el2013accurate}, airborne system~\cite{wang2019joint} and ground-based systems~\cite{luong2020radio}. 


Multistatic radar systems~\cite{wang2021resource} are considered to be a distributed JSC (DJSC) system that consists of multiple spatially diverse JSC nodes located in a large area. Each JSC node is equipped with a sensing function and a communication function. The sensing function is to perform sensing surrounding environments, and it is also able to detect and track  targets for diverse purposes, e.g., estimating channel quality due to signal obstacles and measuring objects. Meanwhile, the communication function is to transmit the information of environment and the target to a centralized controller, i.e., data fusion point, for further processing. In particular, the spatial diversity of JSC nodes provided by the distributed DJSC system allows different aspects of the target to be viewed simultaneously. As a result, the DJSC system has several advantages compared with the monostatic radar systems. The first advantage is that the DJSC system can collect sensing data from a large area. The second advantage is that  the DJSC is able to significantly improve the target detection and tracking performance. This is because of that the spreading of the JSC nodes geometry throughout the surveillance area increases the coverage. As such, the target is likely to be physically close to the JSC nodes during its moving, and thus attaining a higher signal-to-noise ratio (SNR). Moreover, as presented in~\cite{godrich2011power} and \cite{wang2020constrained}, the mean-square error (MSE) of the target localization estimation is inversely proportional to the number of active JSC nodes. As such, the DJSC system with multiple JSC nodes can  improve significantly the localization accuracy. 

However, to exploit the aforementioned advantages, the resource allocation in the DJSC system needs to be addressed. In particular, to increase the surveillance area coverage as well as the localization accuracy, the DJSC system typically deploys a number of JSC nodes. Moreover, each JSC node requires a certain amount of bandwidth for the sensing and communication functions. Due to the fact that the radio resource congestion raises with the rapid growth of IoT devices, the key issue is how to allocate the bandwidth to the sensing and communication functions of the JSC nodes to maximize the efficiency, i.e., maximize the sensing performance and communication performance. Moreover,  
the bandwidth allocation needs to guarantee the fairness among the JSC nodes. The problem is, in fact, challenging since there is always a conflict between the efficiency and the fairness.  

To the best knowledge of authors, this is the first work that addresses both the efficiency and the fairness of the bandwidth allocation in the DJSC system. In particular, we consider a DJSC system that consists of multiple JSC nodes. Each JSC node as a JSC user is equipped with a sensing function and a communication function. Each user uses its sensing function to sense its surrounding environment, detect and track a common target. The user uses its communication function to transmit the information about the target and the surrounding environments to a base station (BS) for further processing. As a case study, we assume that the sensing function performs the target tracking, and thus we use the estimation rate as the sensing performance. Meanwhile, the communication rate is used as the communication performance. The estimation rate and communication rate are directly proportional to the allocated bandwidth. Thus, the BS needs to perform the bandwidth allocation to the sensing functions and communication functions of the JSC users to maximize the total sensing performance, communication performance, and fairness over JSC nodes.


The main contributions of this paper are summarized as follows.
\begin{enumerate}
\item We formulate a bandwidth allocation problem for the DJSC system that aims to optimize bandwidth allocation to the sensing and communication functions of the JSC users. The objective is to maximize the sum of sensing performance, i.e., estimation rate, communication performance, i.e., communication data rate, and fairness of the JSC users. Such an objective design aims to maximize the allocation efficiency and to benefit the spatial diversity of the DJSC system. In particular, to benefit the spatial diversity, we introduce a max-min fairness metric in the objective function which gives a higher relative priority to the users with lower estimation and communication rates. 
\item The optimization problem that aims to maximize the allocation efficiency and fairness is non-convex and difficult to be solved. For this, we propose a fully polynomial time approximation algorithm, namely $\fptas$, that is able to find a near-optimal solution. We then prove that the proposed algorithm can guarantee a near-optimal solution with an accuracy of $\epsilon$. In particular, for any arbitrarily small value of $\epsilon\in (0,1)$, the solution obtained by the proposed algorithm is always at least $1-\epsilon$ times the optimal value. 
\item We further propose to use a heuristic algorithm that performs the bandwidth allocation to the JSC users in a greedy manner. The $\greedy$ algorithm has a linear complexity of $O(N)$, and thus it can be a more suitable solution for the DJSC systems when the short execution time is required. 
\item We provide simulation results to demonstrate the effectiveness of the proposed algorithms. For this, we introduce the active-set sequential quadratic programming algorithm~\cite{ferreau2014qpoases}, namely $\sqp$, that is known as the currently best algorithm for solving non-linear optimization. The simulation results show that our proposed algorithms, i.e., $\fptas$ and $\greedy$, outperforms the $\sqp$ algorithm in terms of system performance, i.e., the estimation rate, communication rate, and fairness, and execution time. 
\end{enumerate}

The rest of the paper is organized as follows. We discuss relevant works in Section~\ref{sec:related-work}. In Section~\ref{sec:system_model_problem}, we describe the DJSC system and formulate the optimization problem. We present the approximation algorithm in Section~\ref{sec:approximation} and the greedy heuristic algorithm in Section~\ref{sec:heuristic}. In Section~\ref{sec:num_result}, we provide and discuss numerical results to verify the effectiveness and improvement of the proposed algorithms. Section~\ref{sec:conclusion} concludes this paper.

\section{Related work}
\label{sec:related-work}
Despite of the fact that the DJSC systems own several promising advantages and that the bandwidth allocation is a major issue, the resource management works in the DJSC systems have not been well investigated. In particular, the authors in \cite{DeligiannisPLC17} consider a DJSC system which is divided into multiple radar sensor clusters. Then, the work addresses the power control of radar sensor clusters to minimize their transmit power while satisfying a certain detection criterion. A game theory is adopted to model the power control strategies of the clusters. Activating all the JSC nodes in the DJSC systems can cause high cost, e.g., bandwidth and power. Thus, the work in \cite{godrich2011power} aims to select a subset of JSC nodes to minimize the total cost while guaranteeing a localization accuracy. Different from~\cite{godrich2011power}, the work in~\cite{wang2020constrained} considers UAVs to be JSC nodes, and then addresses the joint UAV location, user association, and UAV transmission power control problem to maximize the total network utility, i.e., the total data rate, under the constraint of localization accuracy. In \cite{WangL19} the authors investigate joint power allocation for radar and communication systems. The radar system and communication system share the same bandwidth, and the objective is to maximize the performance of one system with a constraint on the throughput for the other system. The authors in~\cite{garcia2014resource} and \cite{li2020adaptive} address the join power and bandwidth allocation problem for the DJSC systems. In particular, the problem is to determine the transmit power and bandwidth of JSC nodes to minimize the lower-bounds on the MSE of localization accuracy. In \cite{ZhangXSZF19}, the strategy of integrating power and bandwidth allocation to the radar sensor selection problem is proposed for the joint multi-target tracking and detection in a distributed MIMO sensor system. 

It should be emphasized in the most of the aforementioned works, efficiency and fairness of resource allocation approaches are not considered, which are very important, especially to the DJSC systems due to the following reasons. First, the DJSC system deploys multiple JSC nodes that can consume a large amount of bandwidth. Meanwhile, the spectrum congestion raises with the growth of IoT devices. Thus, it is necessary to design efficient bandwidth allocation algorithms. Second, to benefit the spatial diversity and to significantly improve the performance of the sensing functions of the DJSC system, the different JSC nodes should be guaranteed to receive adequate bandwidth resources. In other words, the bandwidth allocation algorithms need to be designed to fairness. Due to the conflict of the two criteria, i.e., the efficiency and fairness, most studies have been carried out in past decades on the problem with a single criterion of maximizing efficiency, i.e., throughput in the network, (e.g., \cite{ModianoSZ06,Srinivasan10,ChiWLT18}), or of achieving fairness (see \cite{YinZHFL09} and the surveys \cite{OgryczakLNP14,ShiPON14}). It is very challenging to design a resource allocation scheme to achieve both the efficiency and fairness. In particular, allocating the resources to optimize the utility of worst-off agents may lead to a large worsening of the overall efficiency of the network. Literature shows two main approaches to tackle this issue. The first approach is to model the problem as a two-criteria optimization that simultaneously maximizes both the fairness and efficiency  \cite{Ogryczak2008}. Another direction is to study the trade-off between fairness and efficiency from which one can look for allocations that satisfy these properties only to some extent  \cite{TangWL04,DannaMS12,Joe-WongSLC13,Song16}. Beside the two arformentioned approaches, proportional fairness defined based on utility percentage can be also used to measure the trade-off between fairness and efficiency \cite{kelly98,LiPY08,Mustika15,Abdel14}.

\section{System Model and Problem Formulation}
\label{sec:system_model_problem}

\begin{figure*}[h!]
 \centering
\includegraphics[scale=0.6]{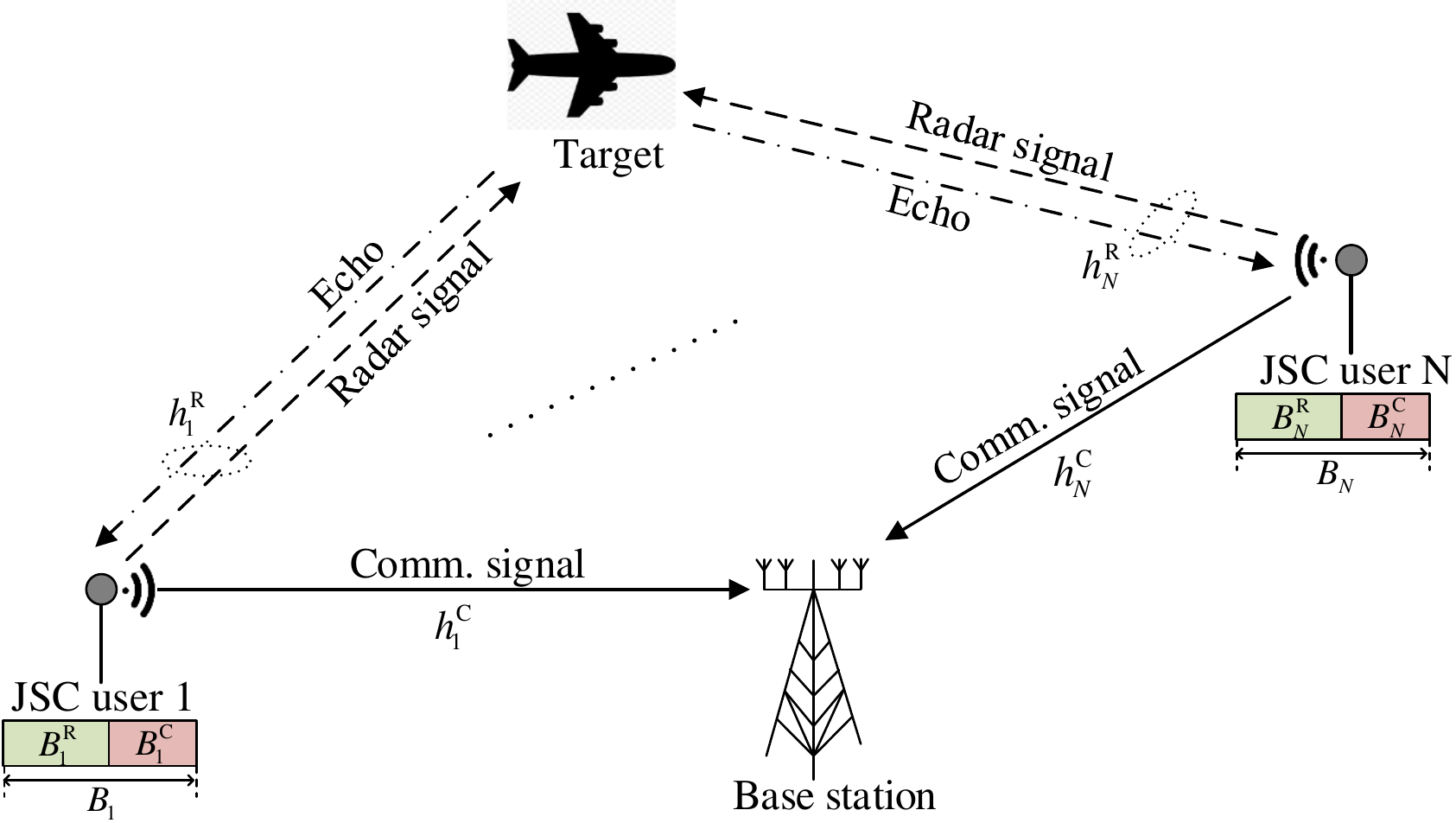}
 \caption{\small A DJSC system with a base station as a data fusion.}
 \label{IRS_model}
\end{figure*}

\subsection{System Model}
We consider a system model as shown in Fig.~\ref{IRS_model} that includes a set $\mathcal{N}$ of $N$ JSC users, where  $\mathcal{N}=\{1,\ldots, N\}$. Each user is equipped with a sensing function and a communication function. In particular, the sensing function acts as an individual radar that is used by the user to detect and track a target. For convenience, we will also use the term of radar in the rest of the paper. The information about the target (target velocity and range) received at the radar receivers of the users is collected at a centralized entity, i.e., a base station (BS), for further processing. To transmit their data to the BS, the users use the communication functions through wireless links. Here, the data of each user includes not only the information of the target but also other sensing data, e.g., of weather condition. As a centralized entity, the BS schedules the sensing function and the communication function for each user in each time slot of a time frame. In particular, let $F$ denote the number of time slots in the time frame, and in time slot $j, 1 \leq j \leq F$, the BS can schedule the sensing function or the communication function for user $i, 1 \leq i \leq N$. Note that in the same time slot, the users may not perform the same function. For example, in time slot $j$, user $i$ can be scheduled to perform the sensing function, and user $k\neq i$ performs the communication function (see Fig.~\ref{IRS_model}).~Moreover, in different time slots of a time frame, the user can perform different functions. For example, in time slot $j+1$, user $i$ can be scheduled to perform the communication function (see Fig.~\ref{IRS_model}). The user is scheduled to perform the same function, i.e., sensing or communication, in the same time slots of the next time frames. For synchronization implementation simplicity, TDMA such as round-robin scheme can be adopted for the function scheduling. To remove the co-channel interference, in each time slot, the BS uses the frequency division multiple access (FDMA) that allocates bandwidth $B$ to the JSC users. In particular, in each time slot $j, 1 \leq j \leq F$, the bandwidth allocation profile for $N$ users is $\textbf{B}_j=(B_{1,j},\ldots, B_{N,j})$, where $B_{i,j} >0$ denotes the portion of bandwidth $B$ allocated to JSC user $i$ in time slot $j$. We have $\sum_{i=1}^NB_{i,j}=B$. In the time slot, each user $i$ can use the allocated bandwidth to perform either the sensing function or the communication function.

Since the total bandwidth $B$ is the same for every time slot, it suffices to consider  the optimization problem within one time slot.  As such, we can omit the time slot index $j$ from the notation to simplify the presentation. In the case that user $i$ is scheduled to perform the communication function, the user uses bandwidth $B_{i}$ to transmit its data to the BS with a data rate given by 

\begin{equation}
g_{i}(B_{i})\triangleq B^{\text{C}}_{i}\cdot \log_2\Big{(} 1+ \frac{||h_i^{\text{C}}||^2\cdot P_i^{r,\text{C}}}{k_B\cdot T_{\text{temp}} \cdot B^{\text{C}}_{i}}  \Big{)},
\end{equation}
where $k_B$ is the Boltzmann constant, $T_{\text{temp}}$ is the absolute temperature, $h_i^{\text{C}}$ is the communication channel gain from user $i$ to the BS, and $P_i^{r,\text{C}}$ is the received power at BS. In particular, $P_i^{r,\text{C}}$ is determined as $P_i^{r,\text{C}} = \frac{P_i^{\text{C}} G_{i,T}^{\text{C}}  G_{i,B}^{\text{C}}}{(4\pi)^2 (d_i^{\text{C}})^2(f_c)^2}$, where $P_{i}^{\text{C}}$ is the communication transmit power of user $i$, $G_{i,T}^{\text{C}}$ is the communication antenna gain of the user, $G_{i,B}^{\text{C}}$ is the receiving antenna gain of the BS, $d_i^{\text{C}}$ is the distance between the user and the BS, and $f_c$ is the carrier frequency. 

In the case that user $i$ performs the sensing function, the user uses bandwidth $B_{i}$ to transmit radar pulses to the target. Let $T_{\text{p}}$ denote the pulse width, and the pulse repetition interval of the radar signal is $T_{\text{pri}}=T_{\text{p}}/\delta$, where $\delta$ is the radar duty factor, i.e., the percentage of the time the radar pulse emits. After receiving the echo from the target, the radar receiver detects and determines parameters such as velocity and range of the target. To measure the sensing performance, we use the estimation rate~\cite{luong2020radio}, \cite{chiriyath2017radar}. The estimation rate is a metric similar to the communication rate that provides a measure of the information about the target. The estimation rate achieved by the JSC user is higher meaning that the amount of information about the target is higher. 

Due to the target tracking, the sensing function of the user has some knowledge of the target, i.e., based on the observations in prior time slots, with an amount of fluctuation, and this fluctuation is called process noise. In particular, we consider the range estimation of the sensing function, and thus the process noise for the range fluctuation is a delay fluctuation that is denoted by $\eta_{\text{proc}}$ with variance of $\sigma^2_{\text{proc}}$. Then, the estimation rate of the sensing function can be determined by~\cite{chiriyath2017radar}
\begin{equation}
f_{i}(B_{i})\triangleq \frac{1}{2T_{\text{pri}}} \log_2\Big{(}   1+ \frac{\kappa \cdot  ||h_i^{\text{R}}||^2 \cdot  P_i^{r,\text{R}}\cdot  B_{i} }{k_B\cdot T_{\text{temp}}} \Big{)},
\end{equation}
 where $P_i^{r,\text{R}}$ is the power received at the sensing receiver of user $i$, $h_i^{\text{R}}$ is the roundtrip channel gain from the JSC user to the target, and $\kappa$ is defined as $\kappa=8\pi^2\sigma^2_{\text{proc}}\gamma^2$, where $\gamma^2=(2\pi)^2/12$. In particular, $P_i^{r,\text{R}}$ is defined as $P_i^{r,\text{R}} = \frac{P_i^{\text{R}} (G_i^{\text{R}})^2 \sigma_{\text{cross}}}{(4\pi)^3 (d_i^{\text{R}})^4(f_c)^2}$, where $P_i^{r,\text{R}}$ is the radar transmit power of user $i$, $G_i^{\text{R}}$ is the radar antenna gain, $d_i^{\text{R}}$ is the distance between the user and the target, and $\sigma_{\text{cross}}$ is the target cross section.


As $B_i$ is allocated to user $i$, the utility achieved by user $i$ is defined as 
\begin{equation}
\label{utility_user_i}
u_{i}({B_i}) \triangleq f_{i}(B_{i}) + g_{i}(B_{i}).
\end{equation}
It can be seen from (\ref{utility_user_i}) that $u_i$ is an increasing function of $B_i$. Here, we consider a general scenario in which there are a lower bound of $\underline \xi_i$ and an upper bound $\overline \xi_i$ on the amount of bandwidth $B_i$ allocated to user $i$. This is due to the fact that in order to make the sensing and communication functions working properly, each user may need to receive at least some fixed amount of bandwidth.
The presence of upper bounds is actually a situation in which there is an upper bound on the amount of bandwidth that a user can receive during peak hours (a.k.a access rate), or there is a budget for each user. Without loss of generality, we consider that $\sum_{i\in\cN} \underline\xi_i \le B$ and $\sum_{i\in\cN} \overline\xi_i > B$. Otherwise, the problem is trivial to solve. 
Also, we consider that $u_i(\underline\xi_i)\ge 1$ for every $i$, without changing optimal solutions of the optimization problem.



\subsection{Optimization Problem}
For easy of presentation, for each user $i\in\mathcal{N}$, we introduce two new variables as follows:
\[
\tau_i=\frac{||h_i^{\text{C}}||^2\cdot P_i^{r,\text{C}}}{k_B\cdot T_{\text{temp}} } \quad \text{and}\quad \nu_i=\kappa  \cdot   \frac{ ||h_i^{\text{R}}||^2 \cdot  P_i^{r,\text{R}}}{k_B\cdot T_{\text{temp}}}.
\]
Moreover, we denote $x_i$ as a variable indicating the amount of bandwidth assigned to $i$. Hence, the data rate and estimation rate are rewritten as follows:
\[
g_i(x)= x_i\cdot \log_2\left(1+\dfrac{\tau_i}{x_i}\right), \,\,  f_i(x_i)= \frac{1}{2T_{\text{pri}}}\cdot \log_2\left(1+\nu_i x_i\right).
\]
Correspondingly, the utility function of user $i$ is defined as the total estimation rate and communication rate as follows:
\[
u_i(x_i) = f_i(x_i) + g_i(x_i).
\]
Then, a vector $\bx=(x_1,\ldots,x_N)\in[0,B]^N$ is said to be a feasible assignment (or solution) if $\sum_{i\in\cN}x_i\le B$ and $x_i\in[\underline \xi_i,\overline \xi_i]$ holds for every $i\in\cN$. Let  $\bu =(u_1,\ldots,u_N)\in\mathbb{R}_+^N$ denote the utility vector corresponding to a feasible assignment $\bx=(x_1,\ldots,x_N)$. As mentioned earlier, our work aims to maximize the efficiency and fairness of the bandwidth assignment. In particular, to measure the efficiency of an assignment $\bx$, we use the $\ell_p$-norm $(p \ge 1)$, which is defined as 
\begin{equation}
\cF_p(\bx)=||\bu||_p=\left ( \sum_{i\in\cN} u_i^p(x_i) \right)^{1/p}.
\label{norm_p}
\end{equation}
To measure the fairness of the assignment, we use the fairness that is defined as  
 \[
\cF_{\min}(\bx)= ||\bu||_{-\infty}=\min_{i\in\mathcal{N}}\{u_i(x_i)\}.
 \]
$\cF_{\min}(\bx)$ is also called the egalitarian welfare of an assignment $\bx$, and the maximum egalitarian social welfare is said to be max-min fair. 
As a centralized entity of the DJSC system, the BS aims to maximize both the efficiency and the fairness of an assignment of bandwidth to users, under the bandwidth constraint. Thus, the optimization problem of the BS, namely \textsc{FEO}\footnote{FEO stands for fair and efficiency optimization.}, is to find a feasible bandwidth allocation $\bx$ to maximize the efficiency and fairness as follows:
\begin{align}
\textsc{FEO}\quad &~  \max_{\bx\in\mathbb{R}_{\ge 0}^{N}}\quad \cF(\bx)= \alpha\cdot\cF_p(\bx)+(1-\alpha)\cdot \cF_{\min}(\bx)  \quad {}&\label{FEO-obj}\\
\text{s.t.} \quad  &~ {\sum}_{i\in\cN} x_{i}\le B,\label{FEO-const1}\\
\text{} \quad  &~ x_i\in[\underline \xi_i,\overline \xi_i],\, i\in\cN, 
\label{FEO-const2}
\end{align}   
where $p,\alpha$ are the scalar parameters $p\ge 1$ and $\alpha\in[0,1]$. The meaning of the value of $\alpha$ is as follows. For $\alpha=0$, the value of $\cF$ is the utility of the worst-off users in the allocation, while for $\alpha=1$, the value of $\cF$ is the sum of the $p$ powers of the utility of users. For $\alpha$ that is strictly between $0$ and $1$, the value of $\cF$ is a convex combination of these classical objectives. Using such an objective function $\cF$ is one of the well-known methods that can help to balance the trade-off between the efficiency and the fairness. In general, the objective function $\cF$ is neither convex nor convave\footnote{The proof of the nonconvexity of $\cF_p$ is given in Appendix.}, leading to a non-convex optimization problem. Hence, there is no standard algorithm for exactly solving such a problem. Therefore, we develop a polynomial time algorithm to determine a near-optimal solution. In fact, for an arbitrary small $\epsilon>0$, the algorithm can theoretically guarantee the value of its found solution $\bx$ to be within a factor of $1-\epsilon$ of the optimal value. The running time of the algorithm is a polynomial in the input size and in $\frac{1}{\epsilon}$. Here, the input size $\cI$ is the total size in binary encoding of the parameters needed to describe $\textsc{FEO}$. These parameters include $N,B,\kappa, k_B, T_{\text{temp}},T_{\text{pri}},h_i^{\text{C}}, h_i^{\text{R}}, P_i^{r,\text{C}} P_i^{r,\text{R}}, \underline \xi_i, \overline \xi_i$, for all $i\in \cN$.

\subsection{The Efficiency-Fairness Trade-off}
Our work aims to maximize the efficiency and the fairness of the bandwidth allocation in the DJSC system. However, there is a trade-off between the efficiency and fairness in the sense that maximizing the efficiency of an allocation might lead to a reduction in its fairness (see an example given in Appendix~\ref{trade-off}). 
In particular, given a certain value of $\alpha$, what the efficiency loss might be, and what the fairness loss might be. To present the trade-off between the efficiency and the fairness, we adapt the work in \cite{Bertsimas2012OnTE} which introduces the concepts of \textit{price of fairness} and \textit{price of efficiency}. For $\alpha\in[0,1]$, let $\bx_{\alpha}$ be an optimal solution to FEO with respect to $\alpha$. We also call such a solution an $\alpha$-allocation. The efficiency loss is the difference between the maximum system efficiency and the efficiency under the fair scheme, that is, $\cF_p(\bx_1)-\cF_p(\bx_{\alpha})$. The price of fairness with respect to $\alpha$ is denoted by $\pof(\alpha)$ and defined as
\[
\pof(\alpha) = \dfrac{\cF_p(\bx_1)-\cF_p(\bx_{\alpha})}{\cF_p(\bx_1)}.
\]
The price of fairness $\pof(\alpha)$ is a nonnegative and less than one, and it measures the percentage of efficiency loss of an $\alpha$-allocation.

Similarly, we can define the fairness loss and the price of efficiency. The fairness loss is the difference between the fairness metric evaluated with the max-min fair allocation and that evaluated with the $\alpha$-fair allocation, i.e., $\cF_{\min}(\bx_0)-\cF_{\min}(\bx_{\alpha})$. Then, we can consider the fairness loss relative to the maximum value of the fairness metric to be the price of efficiency that is denoted by $\poe$ and defined as
\[
\poe(\alpha) = \dfrac{\cF_{\min}(\bx_0)-\cF_{\min}(\bx_{\alpha})}{\cF_{\min}(\bx_0)}.
\]
The price of efficiency $\poe(\alpha)$ can be interpreted as the percentage loss in the minimum utility guarantee compared to the maximum minimum utility guarantee.

In general, both $\pof(\alpha)$ and $\poe(\alpha)$ are functions of $\alpha$ over the domain $[0,1]$. In general, the best value of $\alpha$ to balance efficiency and fairness can be explained through experiments, which is discussed in Section~\ref{sec:num_result}.


\section{An Approximation Algorithm}
\label{sec:approximation}

In this section, we aim at solving the FEO problem defined by (\ref{FEO-obj})-(\ref{FEO-const2}), by designing an approximation algorithm, namely $\fptas$. The result is stated in the following theorem:
\begin{theorem} 
\label{th:fptas}
For arbitrary small constant $\delta\in (0,1)$, there is an algorithm for finding a solution $\bx$ to the \textsc{FEO} problem such that the objective value $\cF(\bx)$ is within a factor of $1-\delta$ of the optimal value. The execution time of the algorithm is polynomial in $\frac{1}{\delta}$ and in the input size.
\end{theorem}


The remaining of this section is devoted to presenting the proof of Theorem~\ref{th:fptas}. 

First, we fix $p\in\mathbb{N}_{\ge 1}$ and an accuracy $\delta\in(0,1)$. 
Let $\delta=6\epsilon$. Let $\bx^*$ be an optimal solution of \textsc{FEO}, and denote  $\cF(\bx^*)=\opt$. We show how to  find a solution $\bx$ with $\cF(\bx)\ge (1-\delta)\cdot \cF(\bx^*)$, by using the algorithm $\fptas$. A formal description can be found in Algorithm~\ref{main-alg}. At a high level, it can be divided into three phases as follows.
 
\begin{itemize}
\item Phase I: Transforming \textsc{FEO} into a parameterized problem, in which the second term of the objective, namely $\cF_{\min}(\bx)$, is put into constraint, using a parameter $\phi$. The resulting problem, denoted by $\textsc{FEO}(\phi)$, is then approximately solved via solving a sequence of problems of suitable fixed values of $\phi$ (see Section~\ref{sec:parameterizing}).
\item Phase II: Solving $\textsc{FEO}(\phi)$ for fixed values of $\phi$ to optimality.  Particularly, we aim to find near-optimal  solutions to such problems, within an accuracy  depending solely on $\epsilon$ (see Section~\ref{sec:solving-FEO}).
\item Phase III: Selecting the solution with the maximum value among the solutions obtained in Phase II. 
\end{itemize}

\begin{algorithm}[!htb]
	\caption{{\fptas}} 
	\label{main-alg}
\begin{algorithmic}[1]
\Require $\{f_i,g_i,\overline\xi_i,\underline \xi_i\}_{i\in\cN}$, and accuracy $\delta>0$
\Ensure A solution $\bx$ with $\cF(\bx)\ge (1-\delta)\cdot \opt$ 
\State $\epsilon \leftarrow \delta/6$;
\State Solve (\ref{phi}) to obtain a value $\overline \phi$, using bisection method
\State $\cS\leftarrow\{(1+\epsilon)^h|~h=0,1,\ldots,\left\lfloor \log_{1+\epsilon}\overline\phi\right\rfloor\}$
\For{$\phi\in \cS$}
   \State Solve $\textsc{FEO}(\phi)$, defined in (\ref{obj})-(\ref{bound-const}), using Algorithm~\ref{fptas}
   \State Let $\bx^{\phi}$ be the obtained solution
\EndFor
\State $\bx\leftarrow\arg \max\{\cF(\bx^{\phi})|~ \bx^{\phi},~\phi\in \cS\}$
\State \Return $\bx$
\end{algorithmic}
\end{algorithm}


In what follows, we will explain in details the phases of the algorithm $\fptas$. Phase I is discussed in Section~\ref{sec:parameterizing}.  At the heart of $\fptas$ is Algorithm~\ref{fptas} for solving $\textsc{FEO}(\phi)$ in Phase II, which is presented in Section~\ref{sec:solving-FEO}. Finally, in Section~\ref{sec:analysis} we will analyze the performance and complexity of $\fptas$, which complete the proof of Theorem~\ref{th:fptas}. To be more focused, some results are stated without proofs, but they can be found in Appendix. 

\subsection{Phase I$-$Parameterization}
\label{sec:parameterizing}
To solve \textsc{FEO}, the idea is introduce a constraint related to $\cF_{\min}$ in the FEO problem. In particular, we consider the $\phi$-parametrized \textsc{FEO} as follows:
\begin{align}
\textsc{FEO}(\phi)\quad &~  \max_{\bx\in\mathbb{R}_{\ge 0}^{N}}\quad \alpha\cdot\cF_p(\bx)+(1-\alpha)\cdot \phi   \quad {}&\label{obj}\\
\text{s.t.} \quad  &~ {\sum}_{i\in\cN} x_{i}\le B,\label{capacity}\\
\text{} \quad  &~ \cF_{\min}(\bx) \ge \phi, \label{phi-const}\\
\text{} \quad  &~ x_i\in[\underline \xi_i,\overline \xi_i],\, i\in\cN, \label{bound-const}
\end{align}   
where $\phi\in\mathbb{R}_{>0}$. Since the utility functions are increasing, we can assume that the equality in (\ref{capacity}) is replaced by a "$\leq$".  It is observed that $\textsc{FEO}(\phi)$ may not be feasible, as there may not exist any feasible solution satisfying the constraints in (\ref{capacity}), (\ref{phi-const}) and (\ref{bound-const}), for some value $\phi$. To deal with this issue, we may restrict the value of $\phi$ to the range $[\phi_1^*,\phi_2^*]$, where 
\begin{align}
\nonumber\phi_1^*=&\,\min_{\pmb{ \underline\xi}\le\bx\le \pmb{ \overline\xi}} \cF_{\min}(\bx)=\cF_{\min} (\pmb{ \underline\xi})=\min_{i\in\cN} \{u_i(\underline \xi_i)\},\\ 
\label{phi}\phi_2^*=&\,\max_{\pmb{ \underline\xi}\le\bx\le \pmb{ \overline\xi}}\{ \cF_{\min}(\bx):{\sum}_{i\in\cN} x_i\le B\}=\phi^*. 
\end{align}

\begin{claim}
\label{cl:phi}
We can compute an approximate value of $\phi^*$, says $\overline \phi$, to within any accuracy $\delta'>0$. Furthermore, by choosing a suitable value of $\delta'$ as a function of $\epsilon$, we can guarantee that $\overline \phi\ge (1-\epsilon)\cdot \phi^*$. 
\end{claim}
 We denote by $\opt(\phi)$ as the optimal value of  $\textsc{FEO}(\phi)$, with respect to a given parameter $\phi$. From Claim~\ref{cl:phi}, we can prove that
\begin{equation}
\label{eq:opt-relation}
\max_{1\le\phi\le\overline \phi} \{\opt(\phi)\}\ge (1-\epsilon)\cdot\max_{1\le\phi\le \phi^*} \{\opt(\phi)\} = (1-\epsilon)\cdot\opt.
\end{equation}

We now find $\phi\in[1,\overline \phi]$ such that  $\opt(\phi)$ is maximized. For this, we consider two subproblems. The first one is to determine the value of $\textsc{FEO}(\phi)$ with a fixed value of $\phi$. After the first subproblem is solved, the second subproblem is to determine $\phi$ that maximizes $\opt(\phi)$. Note that the second subproblem is nonconvex (as discussed earlier). Hence, we propose a discrete method to find its approximate solutions. 
For this purpose, we partition the interval $[1,\overline \phi]$ into disjoint intervals as follows.
\[
[1,\overline \phi]=\bigcup_{h=0}^{H-1} [(1+\epsilon)^h,(1+\epsilon)^{h+1}) \bigcup [(1+\epsilon)^H,\overline\phi),
\]
where $H=\left\lfloor \log_{1+\epsilon}\overline\phi\right\rfloor$. Let $\cS=\{(1+\epsilon)^h|~h=0,1,\ldots,H\}$. Then, we have the following lemma:
\begin{lemma}
\label{lem:1}
It holds that 
\[
\max_{\phi\in\cS} \{\opt(\phi)\} \ge (1-\epsilon)\cdot \max_{1\le\phi\le\overline \phi} \{\opt(\phi)\}.
\]
\end{lemma}
\begin{proof}
Suppose that $\tilde{\phi}=\arg\max\{\opt(\phi)|~1\le\phi\le\overline \phi\}$. Then, it must hold that $\tilde{\phi}$ belongs to some intervals in $[(1+\epsilon)^h,(1+\epsilon)^{h+1})$ for some values of $h\in\{0,1,\ldots,H\}$. It is clear that every feasible solution to $\textsc{FEO}(\tilde{\phi})$ is also feasible to $\textsc{FEO}((1+\epsilon)^h)$. Let $\bx$ be an optimal solution to $\textsc{FEO}(\tilde{\phi})$. Then, $\bx$ is also feasible to $\textsc{FEO}((1+\epsilon)^h)$, and we have 
\begin{align*}
\opt((1+\epsilon)^h)&~=\alpha\cdot\cF_p(\bx)+(1-\alpha)\cdot (1+\epsilon)^h\\
&~\ge\alpha\cdot\cF_p(\bx)+(1-\alpha)\cdot \frac{\tilde{\phi}}{1+\epsilon}\\
&~\ge \frac{1}{1+\epsilon} (\alpha\cdot\cF_p(\bx)+(1-\alpha)\cdot \tilde{\phi}).
\end{align*}
By the definition of $\tilde{\phi}$ and the fact that $\epsilon>0$, the proof of Lemma \ref{lem:1} is completed. 
\end{proof}

Lemma~\ref{lem:1} together with (\ref{eq:opt-relation}) imply that
\begin{equation}
\label{eq:core}
\max_{\phi\in\cS} \{\opt(\phi)\} \ge  (1-\epsilon)\cdot\opt,
\end{equation}
and thus solving $\textsc{FEO}$ is now to find the solution to a series of problems $\textsc{FEO}(\phi)$ for $\phi\in\cS$. We propose an approximation scheme, $\fptas$, for solving such parameterized problems in the next section.  

\subsection{Phase II$-$Solving $\textsc{FEO}(\phi)$}
\label{sec:solving-FEO}
By fixing $h\in\{0,1,\ldots, H\}$ and considering that $\phi=(1+\epsilon)^h$, we develop Algorithm~\ref{fptas} for solving $\textsc{FEO}(\phi)$. The key idea is to discrerize $\textsc{FEO}(\phi)$  as a {\em multiple-choice Knapsack problem} (\textsc{MCKP}) \cite{han04}, which can be then well approximated by adapting dynamic programming techniques. In what follows we assume w.l.o.g that $\alpha\not=0$.

To simplify the presentation, we ignore the constant term $(1-\alpha)\phi$ in the objective function of $\textsc{FEO}(\phi)$, as this does not change optimal solutions to the original problem.  
We denote the problem, obtained from $\textsc{FEO}(\phi)$ by replacing $\cF(\bx)$ by $\cF_p(\bx)$, as $\overline{\textsc{FEO}}(\phi)$. Algorithm~\ref{fptas} essentially consists of two main steps:
\begin{itemize}
\item {\em Discretization: } Relaxing $\overline{\textsc{FEO}}(\phi)$ as a binary optimization problem with only $0$-$1$ variables, which is called $\textsc{MCKP}(\phi)$.
\item {\em Solving $\textsc{MCKP}(\phi)$: } Running dynamic programming on $\textsc{MCKP}(\phi)$.
\end{itemize}

\subsubsection{Discretization} 
\label{sub:discret}
Let $\bar \bx$ be an optimal solution to $\overline{\textsc{FEO}}(\phi)$. 
First, we derive a {\it positive} lower bound $L$ on $\cF_p(\bar\bx)$, which can be defined as $L\triangleq \cF_p(\pmb{ \underline\xi})$. For each $i\in\cN$, we define 
\begin{equation}
u_i^j\triangleq\frac{\epsilon L}{\sqrt[p]{N}}(1+\epsilon)^j,
\end{equation}
for $j=0,1\ldots,K_i\triangleq\left\lceil\log_{(1+\epsilon)}\frac{\sqrt[p]{N} u_i(\overline\xi_i)}{\epsilon L}\right\rceil$. As $u_i(\cdot)$ is strictly monotone increasing, then $u_i(x)=u_i^j$, for $j=0,1,\ldots,K_i$, has a unique positive root $x_i^j$ in the interval $(0,\overline\xi_i]$. Given any desired accuracy $\epsilon_i>0$, we can approximate this root to within an absolute error of $\epsilon_i$ using the bisection method:
\begin{align}\label{approx-root}
|\tilde x_i^j-x_i^j|\le\epsilon_i,
\end{align}
in time $O(\log_2 \frac{u_i(\overline\xi_i)}{\epsilon_i})$, where $\tilde x_i^j$ denotes the approximate root. Furthermore, one can show the following Lemma (its proof can be found in Appendix). 
\begin{lemma}
\label{claim:epsilon}
By choosing $\epsilon_i$ small enough, e.g., $\epsilon_i:=\frac{\epsilon^2L}{\sqrt[p]{N}}: \left( \frac{\nu_i}{2T_{\text{pri}}}+\frac{\tau_ic_i\sqrt[p]{N}}{\epsilon L}\right)$, for some positive number $c_i$, it follows that $|u_i(\tilde x_i^j)-u_i(x_i^j)|\le\epsilon  u_i(x_i^j)$.
\end{lemma}

Lemma \ref{claim:epsilon} implies that
\begin{align}\label{approx-f}
u_i(\tilde x_i^{j})\le u_i(x)\le \frac{(1+\epsilon)^2}{1-\epsilon}u_i(\tilde x_i^j),
\end{align}
for all $j=0,1,\ldots,K_i-1$ and $x\in[\tilde x_i^{j},\tilde x_i^{j+1}]$. Note that for $x\in[\tilde x_i^j,\tilde x_i^{j+1}]$, we have
\begin{align}
u_i(\tilde x_i^{j})\le u_i(x)\le u_i(\tilde x_i^{j+1}).
\label{u_x_inequal}
\end{align}
By combining ($\ref{u_x_inequal}$) with (\ref{approx-f}), we have 
\begin{align*}
u_i(x)&\le (1+\epsilon)u_i(x_i^{j+1})=(1+\epsilon)(1+\epsilon)^{j+1}\frac{\epsilon L}{\sqrt[p]{N}}\\
&=(1+\epsilon)^2u_i(x_i^j)\le \frac{(1+\epsilon)^2}{1-\epsilon}u_i(\tilde x_i^j).
\end{align*}


The following lemma estimates the execution time required for the above discrete process. Let 
\begin{equation}
\label{eq:omega}
\omega =   \max_{i\in\cN} \left\{\log^2_2 u_i(\overline\xi_i)\cdot \log_2 \left( \frac{\nu_i}{2T_{\text{pri}}}+\tau_ic_i   \right)\right\},
\end{equation}
 where  $c_i=\max\{2,2\log_2 \frac{\tau_i}{\epsilon}\}$. 

\begin{lemma}
\label{cl:discrete}
The discretization can be implemented in time $\tim_{\text{discrete}}=O(\frac{1}{\epsilon}\log^2_2 \frac{1}{\epsilon}\cdot N\cdot \omega)$. 
\end{lemma}


We now show how to relax $\overline{\textsc{FEO}}(\phi)$ as a $\textsc{MCKP}(\phi)$ problem. Recall that in multiple-choice knapsack problem, given a set of $N$ users and a discrete demand set $D_i$ for each user $i$, a profit function\footnote{We use the terms ``profit function" here to differentiate it from ``utility function".} $U_i:D_i\to\mathbb{R}_+$, generates the profit for user $i$ if its demand $d_i\in D_i$ is fulfilled.  
 Given a bound $B$ on the total demand, the objective is to choose a demand $d_i\in D_i$ for each user, so as to maximize the total profit $\sum_{i\in\cN}U_i(d_i)$ subject to $\sum_{i\in\cN}d_i\le B$. Mathematically, we can formulate the problem as a binary linear program:
\begin{align*}
\max \quad &~  {\sum}_{i\in\cN}{\sum}_{k=1}^{|D_i|} U_i(d_{ik})\cdot z_{ik} \quad {}&\\
\text{s.t.} \quad  &~ {\sum}_{i\in\cN}{\sum}_{k=1}^{|D_i|} d_{ik}\cdot z_{ik} \le B, \\
&~ {\sum}_{k=1}^{|D_i|} z_{ik} = 1, \quad \text{for} ~i\in\cN\\
&~ z_{ik}\in\{0,1\}, \quad \text{for} ~i\in\cN,~k=1,\ldots, |D_i|,
\end{align*}   
where $z_{ik}=1$ indicates that the demand $d_{ik}$ is chosen from the set $D_i$ by user $i$.

We can model the discretized version of $\overline{\textsc{FEO}}(\phi)$ as an $\textsc{MCKP}(\phi)$ as follows. For user $i$, we define $D_i$ as the set of $\underline\xi_i,\overline\xi_i$ and roots $\tilde x_i^j$ for which the utility of the user is greater than or equal to $\phi$. We define the profit function of user $i$ by $
U_i(d_i)\triangleq \left(u_i(d_i)\right)^p,
$ 
for every $d_i\in D_i$.

The following lemma gives the relationship between $\overline{\textsc{FEO}}(\phi)$ and its discrete version, $\textsc{MCKP}(\phi)$, in terms of solution.
\begin{lemma}
\label{lem:relation}
Given an optimal solution $\{x_i^*\}_i$ for $\overline{\textsc{FEO}}(\phi)$, there is a feasible solution $\{d_i\}_i$ to $\textsc{MCKP}(\phi)$ such that $(\sum_{i\in\cN}U_i(d_i))^{1/p}\ge (1-4\epsilon)\cdot\cF_p(\bar\bx)$. 
\end{lemma}
\begin{proof}
We have $(\cF_p(\bar\bx))^p={\sum}_{i\in\cN} (u_i(x_i^*))^p.$
 Let $\bar d_i$  be the largest element in $\{\tilde x_i^j\}_j$  such that $\bar d_i\le x_i^*$.  
We define the solution $\{d_i\}_i$ to $\textsc{MCKP}(\phi)$ as 
$
d_i\triangleq\bar d_i.
$
 Note that this gives a solution to $\textsc{MCKP}(\phi)$ with finite utility.
Let $J$ be the set of users in which for each user $i$ in the set, $u_i(x_i^*)\ge \frac{\epsilon L}{\sqrt[p]{N} }$. Then by (\ref{approx-f}), for any $i\in J$, we have
$u_i(\bar d_i)\ge\frac{(1-\epsilon)}{(1+\epsilon)^2}u_i(x_i^*)$. 
On the other hand, 
\[
\sum_{i\not\in J} (u_i(x_i^*))^p\le  \left(\frac{\epsilon L}{\sqrt[p]{N}}\right)^p\cdot N\le \epsilon^p\cdot(\cF_p(\bar\bx))^p,
\]
as $N\ge 1$. 
  It follows that
\begin{align*}
{\sum}_{i\in\cN}U_i(d_i)&\triangleq   \sum_{i\in\cN}(u_i(d_i))^p\ge \sum_{i\in\cN}(u_i(\bar d_i))^p\\
&\ge\frac{(1-\epsilon)^p}{(1+\epsilon)^{2p}}\sum_{i\in J}(u_i(x_i^*))^p\\
&\ge\frac{(1-\epsilon)^p(1-\epsilon^p)}{(1+\epsilon)^{2p}}(\cF_p(\bar\bx))^p\\
&\ge\frac{(1-\epsilon)^{2p}}{(1+\epsilon)^{2p}}(\cF_p(\bar\bx))^p\ge(1-4\epsilon)^p(\cF_p(\bar\bx))^p,
\end{align*} 
for $\epsilon\in(0,1)$ and $p\ge 1$. Therefore,
\begin{align}
\label{eq:factor}
\left({\sum}_{i\in\cN}U_i(d_i)\right)^{1/p}\ge(1-4\epsilon)\cdot \cF_p(\bar\bx),
\end{align} 
and this completes the proof of Lemma~\ref{lem:relation}.~\end{proof}

\begin{algorithm}[!htb]
	\caption{} 
	\label{fptas}
\begin{algorithmic}[1]
\Require $\{f_i,g_i,\overline\xi_i,\underline \xi_i\}_{i\in\cN}$, $\phi$, and accuracy $\delta>0$
\Ensure A solution $\bx^\phi$ with $\cF(\bx^\phi)\ge (1-\delta)\cdot \text{\opt}(\phi)$ 
\LineComment{// Setting the lower bound on $\opt(\phi)$}
      \State $L\leftarrow \left(\sum_{i\in\cN}\left(u_i(\underline\xi_i)\right)^p\right)^{1/p}$
\State $\epsilon \leftarrow \delta/6$;
\LineComment{// Discretization}
\For{$i=1$ to $N$}
   \State $K_{i}\leftarrow\left\lceil\log_{(1+\epsilon)}\frac{\sqrt[p]{N}u_i(\overline\xi_i)}{\epsilon L}\right\rceil$
   \State $\tilde x_i^j \leftarrow$ root of $u_i(x)=\frac{\epsilon L}{\sqrt[p]{N}}(1+\epsilon)^j$, $j\in [K_{i}]\cup\{0\}$
   \State $D_i\triangleq \big\{\tilde x_i^j\big\}_j\cup\{\underline\xi_i,\overline\xi_i\}\bigcap \{x_i:u_i(x_i)\ge \phi\}$
   \State $U_i(d_i)\leftarrow \left(f_i(d_i)+g_i(d_i)\right)^p$ for $d_i\in D_i$
\EndFor
\LineComment{// Approximately solve $\textsc{MCKP}(\phi)$}
\State $\theta \leftarrow \dfrac{\epsilon }{N}\cdot\max_{i,k}\{U_i(d_{ik})\}$
\For{$i=1$ to $N$}
     \State $\widetilde U_i(d_i)\leftarrow \lfloor\frac{U_i(d_i)}{\theta}\rfloor$, for $d_i\in D_i$
\EndFor
\State Apply {\bf DP} with $\{N,N',\{D_i,\widetilde U_i\}_{i\in\cN}\}$ as input to obtain a solution $\{d_i^*\}_{i\in\cN}$
\State \Return $\{d_i^*\}_{i\in\cN}$ as $\bx^\phi$
\end{algorithmic}
\end{algorithm}

\subsubsection{A dynamic program (DP) for $\textsc{MCKP}(\phi)$}
\label{sub:DP}
Given an input $\{D_i,U_i\}_{i\in\cN}$ of $\textsc{MCKP}(\phi)$, for each $i\in \mathcal{N}$, we can scale its utility by $\lfloor\frac{U_i(d_i)}{\theta}\rfloor$, for all $d_i\in D_i$. As shown in \cite{han04}, to speed up DP, one can choose $\theta$  as $\frac{\epsilon Z}{N}$, where $Z$ is the value of an optimal solution to the linear programming (LP) relaxation of $\textsc{MCKP}(\phi)$, in which the $0$-$1$ variables $z_{ik}$ are replaced by continuous ones in the interval $[0,1]$. This LP can be solved in linear time in $N$ (see, e.g., \cite{Dyer1984AnOA}). The dynamic program applied to the $\textsc{MCKP}(\phi)$ with the scaled input can be described Algorithm~\ref{dp}. We denote $\zeta_i(a)$ as the minimal bandwidth of a solution of the subproblem consisting of the classes $D_1,\cdots,D_N$ with total utility equal to $q$. If no solution with utility $q$ exists, we set $\zeta_i(a):=B+1$. Let $k_i=|D_i|$ for $i\in\cN$, and let $N':=\lceil N/\epsilon\rceil$. By the scaling procedure, DP may not be guaranteed to attain an exact optimal solution to $\textsc{MCKP}(\phi)$. Nevertheless, the advantage of the algorithm is that it has a polynomial complexity and provides a near-optimal solution with a bound on the quality, as stated in Lemma~\ref{lem:dp-sol} and Lemma~\ref{lem:complexity-DP} below.

\begin{lemma}[\cite{han04}]
\label{lem:dp-sol}
The value of the solution $\{d_i^*\}_{i\in\cN}$ returned by Algorithm \ref{dp} is at least $1-\epsilon$ times the optimal value of $\textsc{MCKP}(\phi)$.
\end{lemma}

\begin{algorithm}[!htb]
	\caption{{\bf DP}} 
	\label{dp}
\begin{algorithmic}[1]
\Require $N,N',\{D_i,\widetilde U_i\}_{i\in\cN}$
\Ensure A solution  $\{d_i^*\}_{i\in\cN}$
\LineComment{// Initialization step}
\State $\zeta_0(0)\leftarrow 0$
\For{$a\in \{1,\ldots,N'\}$}
   \State $\zeta_0(a)\leftarrow B+1$
\EndFor
\LineComment{// Recursion step}
\For{$i\in\cN$}
     \For{$a\in \{0,1,\ldots,N'\}$}
          \State compute $\zeta_i(a)$ as
\begin{align*}
\min\left\{\begin{array}{ll}
\zeta_{i-1}(a-\widetilde U_{i}(d_{i1}))+  d_{i1}&\text{if }a\ge \widetilde U_{i}(d_{i1})\\
\zeta_{i-1}(a-\widetilde U_{i}(d_{i2}))+  d_{i2}&\text{if }a\ge \widetilde U_{i}(d_{i2})\\
\ldots\ldots\ldots\\
\zeta_{i-1}(a-\widetilde U_{i}(d_{ik}))+  d_{ik_i}&\text{if }a\ge \widetilde U_{i}(d_{ik_i})
\end{array}
\right.
\end{align*}
     \EndFor
\EndFor
\State $t\leftarrow \max\{a|~ \zeta_N (a)\le B\}$
\State \Return Solution $\{d_i^*\}_{i\in\cN}$ corresponding to the value $t$.
\end{algorithmic}
\end{algorithm}

The  complexity of $\textsc{DP}$ is stated in following lemma, whose proof can be found in Appendix.
\begin{lemma}
\label{lem:complexity-DP}
The dynamic programming can be implemented in time $\tim_{\textsc{DP}}=O(\frac{1}{\epsilon^2}\log_2\frac{1}{\epsilon}N^2\max_{i\in\cN}\{\log_2 u_i(\overline\xi_i)\})$. 
\end{lemma}

We conclude this section by providing the complexity of Algorithm~\ref{fptas}. One can see that this complexity is dominated by that of the discrerization process, and of the dynamic programming algorithm. Therefore, from Lemma~\ref{cl:discrete} and Lemma~\ref{lem:complexity-DP}, it follows that the overall complexity of Algorithm~\ref{fptas}, $\tim_{\text{Alg}_2}$, is upper bound by $O(\tim_{\text{discrete}}+ \tim_{\text{DP}})=O(\frac{1}{\epsilon^2}\log^2_2\frac{1}{\epsilon}N^2\omega)$, where $\omega$ is given in (\ref{eq:omega}), due to Lemma~\ref{cl:discrete} and Lemma~\ref{lem:complexity-DP}. 
\begin{lemma}
\label{lem:complexity-alg2}
The Algorithm~\ref{fptas} can be implemented in time $\tim_{\text{Alg}_2}O(\frac{1}{\epsilon^2}\log^2_2\frac{1}{\epsilon}N^2\omega)$. 
\end{lemma}

\subsection{Performance and Complexity analysis}
\label{sec:analysis}
In this section we will analyze the complexity and the performance of Algorithm~\ref{fptas}, and thus complete the proof of Theorem~\ref{th:fptas}.

{\em Performance analysis:} We show that the solution $\bx$ returned by $\fptas$ fulfills the quality bound of $\cF(\bx)\ge (1-6\epsilon)\cdot \opt= (1-\delta)\cdot \opt$, as desired. To accomplish this goal, we first prove that the solution $\{d_i^*\}_{i\in\cN}$ (or $\bx^\phi$) obtained by the DP has the value of at least $1-5\epsilon$ times the optimal value of $\overline{\textsc{FEO}}(\phi)$. Let  $\{d_i\}_{i\in\cN}$ be the solution to $\textsc{MCKP}(\phi)$ defined in Lemma~\ref{lem:relation}, and recall that $\bar \bx$ is the optimal solution to $\overline{\textsc{FEO}}(\phi)$. By Lemma~\ref{lem:dp-sol}, we have that
\begin{align*}
\left({\sum}_{i\in\cN} U_i(d_i^*)\right)^{1/p}\ge&~ (1-\epsilon)^{1/p}  \left({\sum}_{i\in\cN} U_i(d_i)\right)^{1/p}\\
\ge&~ (1-\epsilon)(1-4\epsilon)\cdot\cF_p(\bar\bx)\\
\ge&~ (1-5\epsilon)\cdot\cF_p(\bar\bx),
\end{align*}
 where the second inequality follows from (\ref{eq:factor}) and the fact that $p\ge 1$ and $\epsilon> 0$. 
On the other hand, we have
\begin{align*}
\cF(\bx^\phi)=&~\alpha\left({\sum}_{i\in\cN} U_i(d_i^*)\right)^{1/p} +(1-\alpha)\phi\\
\ge&~ \alpha(1-5\epsilon)\cdot\cF_p(\bar\bx)+(1-\alpha)\phi\\
\ge&~ (1-5\epsilon)\cdot\opt(\phi),
\end{align*}
as $\epsilon<1/5$. 

Since $\bx$ is returned by $\fptas$, it must hold that
\begin{align*}
\cF(\bx)=\max_{\phi\in\cS} \{\cF(\bx^\phi)\}\ge  (1-5\epsilon)\cdot\max_{\phi\in\cS} \{\opt(\phi)\}.
\end{align*}
Finally, using (\ref{eq:core}), we obtain
\begin{equation}
\cF(\bx) \ge  (1-5\epsilon)(1-\epsilon)\cdot\opt \ge (1-6\epsilon)\cdot\opt.
\end{equation}

{\em Complexity analysis:} Since $\delta=6\epsilon$, it suffices to prove that the execution time of $\fptas$ is a polynomial in $\frac{1}{\epsilon}$ and in the input size. 
Indeed, we observe from the previous section that an upper bound on the execution time of $\fptas$ is $O(|S|\cdot \tim_{\text{Alg}_2})$, where $\tim_{\text{Alg}_2}=O(\frac{1}{\epsilon^2}\log^2_2\frac{1}{\epsilon}N^2\omega)$, as shown in Lemma~\ref{lem:complexity-alg2}. We now upper bound the size of $\cS$. Note that $|\cS|= H\le \log_{1+\epsilon}\overline\phi$, and by the definition of $\overline\phi$ we have that $\overline\phi \le \cF_{\min} (\pmb{ \overline\xi}) =\min_{i\in\cN} \{u_i(\overline \xi_i)\}\le \max_{i\in\cN} \{u_i(\overline \xi_i)\}$. Hence, $|\cS|\le \log_2\overline\phi/\log_2(1+\epsilon)\le  \frac{1}{\epsilon}\max_{i\in\cN}\{ \log_2 u_i(\overline \xi_i)\}$, as $0<\epsilon<1$. Finally, it follows that the execution time of $\fptas$ is 
 \begin{equation}
 \label{eq:execution-fptas}
 O\left(\frac{1}{\epsilon^3}\log^2_2\frac{1}{\epsilon}N^2\omega\max_{i\in\cN}\{ \log_2 u_i(\overline \xi_i)\}\right),
 \end{equation}
which, by the definition of $\omega$, is polynomial in $\frac{1}{\epsilon}$ and in the input size.

\section{A Heuristic Algorithm}
\label{sec:heuristic}

The approximation algorithm presented in the previous section can mathematically guarantee that the solution is arbitrarily closed to its optimal value as long as the value of $\epsilon$ is small enough. However, its execution time, i.e., running time, is cubic in the input size and in $\frac{1}{\epsilon}$, and thus is not a {\em strongly polynomial-time algorithm}\footnote{A strongly polynomial algorithm is a  polynomial in the number of users, and is independent of the input size.}. In this section, we propose to use a heuristic algorithm, namely $\greedy$, to solve $\textsc{FEO}$. The advantage of $\greedy$ is that it runs in linear time in $N$. The general idea is to find an allocation $(B_1,\ldots,B_N)$ of bandwidth among users in a greedy manner.  In particular, the $\greedy$ allocates an amount of bandwidth of $\underline \xi_i$ to user $i$ in order to fulfill the constraints in (\ref{bound-const}). The remaining amount of bandwidth, $B'$, is greedily allocated to the users according to ratios of their utility gains. For this, the algorithm first computes, for each user $i$, the ratio $\delta_i=\frac{u_i(\overline \xi_i)-u_i(\underline \xi_i)}{\overline \xi_i-\underline \xi_i}$. The algorithm then assigns to each user $i$ an amount of bandwidth which is proportional to $\delta_i$ with a proportionality coefficient of exactly $B'$.  Algorithm~\ref{greedy} shows steps to implement the $\greedy$ algorithm. It can simply be verified that the $\greedy$ algorithm has a linear complexity of $O(N)$. Surprisingly, as shown in the next section, the $\greedy$ algorithm can significantly improve the performance in terms of fairness and efficiency compared with the baseline algorithm.

\begin{algorithm}[!htb]
	\caption{$\greedy$} 
	\label{greedy}
\begin{algorithmic}[1]
\Require $\{f_i,g_i,\overline\xi_i,\underline \xi_i\}_{i\in\cN}$, a capacity $B$
\Ensure An allocation $(B_1,\ldots,B_N)$
\State Construct a vector $S=\{\delta_i=\frac{u_i(\overline \xi_i)-u_i(\underline \xi_i)}{\overline \xi_i-\underline \xi_i}~|~i\in\cN\}$
 \label{step:seq-const}
\LineComment{// Normalize vector $S$}
 \State $\bar\delta_i \leftarrow \frac{\delta_i}{\sum_{i\in \cN} \delta_i}$, for $i\in\cN$
 \State $\bar S  \leftarrow (\bar\delta_i)_{i\in\cN}$
\LineComment{// Allocate bandwidth to users based on $\bar S$}
\State $B\leftarrow B-\sum_{i\in\cN} B_i$
\If{$B>0$}
   \State $B_{i}\leftarrow \underline \xi_i + \min\{ \overline\xi_i-\underline \xi_i, \bar\delta_i\cdot B  \}$, for $i\in\cN$
\EndIf
\State \Return $(B_1,\ldots,B_N)$ 
\end{algorithmic}
\end{algorithm}

\section{Numerical Results}
\label{sec:num_result}

In this section, we present experimental results to evaluate our two proposed algorithms, i.e., the $\fptas$ and $\greedy$ algorithms. For the comparison purpose, we introduce the active-set sequential quadratic programming ($\sqp$) algorithm~\cite{ferreau2014qpoases}, which is known as the best algorithm for nonlinear programming, as the baseline algorithm.

For the evaluation purpose, we consider a DJSC system in which the number of users is $N\in\{2,3,\ldots,10\}$. To measure the efficiency of the proposed resource allocation algorithms, we use the Euclidean norm $\ell_2$ for $\cF_p(\bx)$, i.e., $p=2$. All the involved channels in the DJSC system are uniformly distributed in the interval $(0.5,1)$. The upper bound is $\overline \xi_i=10^7$, and the lower bound is $\underline \xi_i=10^4$. We implement all the algorithms using Python on an \textsc{Intel} Core i$7$, $1.5$ GHz, with $16$ GB of RAM. Other simulation parameters are given in Table~\ref{table:parameters}, which are also similar to those in~\cite{chiriyath2017radar} and \cite{bliss2014cooperative}. In particular, the accuracy parameter, i.e., $\epsilon$, and the weight parameter, i.e., $\alpha$, are the important parameters, whose values significantly impact the system performance and execution time of the $\fptas$ algorithm. Thus, in the followings, we first evaluate the proposed algorithm, i.e., the $\fptas$ algorithm, by varying $\epsilon$ and $\alpha$, which can help us to select the proper values of $\epsilon$ and $\alpha$. Then, we compare the proposed algorithms and the baseline algorithm. We highlight that the term of efficiency refers to the objective of maximizing the total estimation rate and communication rate over the users. 


\begin{table}[!h]
\caption{Simulation parameters}
\label{table:parameters}
\centering
\begin{tabular}{llc}
\hline\hline
{Parameters} 		& {\em Value} \\ [0.5ex]
\hline
Total bandwidth ($B$)   &  $10^{7}$ Hz\\
Carrier frequency ($f_c$) & $10^8$ Hz \\ 
Communication range ($d^{\text{C}}_i$)        & $10^2 $ m   \\ 
Communication transmit power ($P_i^{\text{C}}$)   & $43$ dBm  \\ 
Gain of communication antenna of JCS users ($G_{i,T}^{\text{C}}$)  & $19$ dB   \\ 
Gain of receiving antenna of the BS ($G_{i,B}^{\text{C}}$)  & $19$ dB   \\ 
Target range ($d^{\text{R}}_i$)      &$5\times 10^3$ m   \\ 
Radar antenna gain ($G_i^{\text{R}}$)  & $30$ dBi   \cite{chiriyath2017radar}  \\ 
Radar transmit power ($P_i^{\text{R}}$)   & $100$ kW \cite{chiriyath2017radar}  \\ 
Target cross section ($\sigma_{\text{cross}}$)  & $10$ m$^2$  \\ 
$\sigma_{\text{proc}}$  & $10^2$ m  \\ 
$k_B$ & $1.38\times10^{-23}$   \\ 
$T_{\text{temp}}$ & $10^3$   \\ 
$T_{\text{pri}}$ & $10^{-5}$   \\ 
\hline
\end{tabular}
\label{table:parameters}
\end{table}

\subsection{Evaluation of the $\fptas$ Algorithm}
\label{sec:evaluation}
\begin{figure}[h!]
  \centering
  \includegraphics[width=0.6\linewidth]{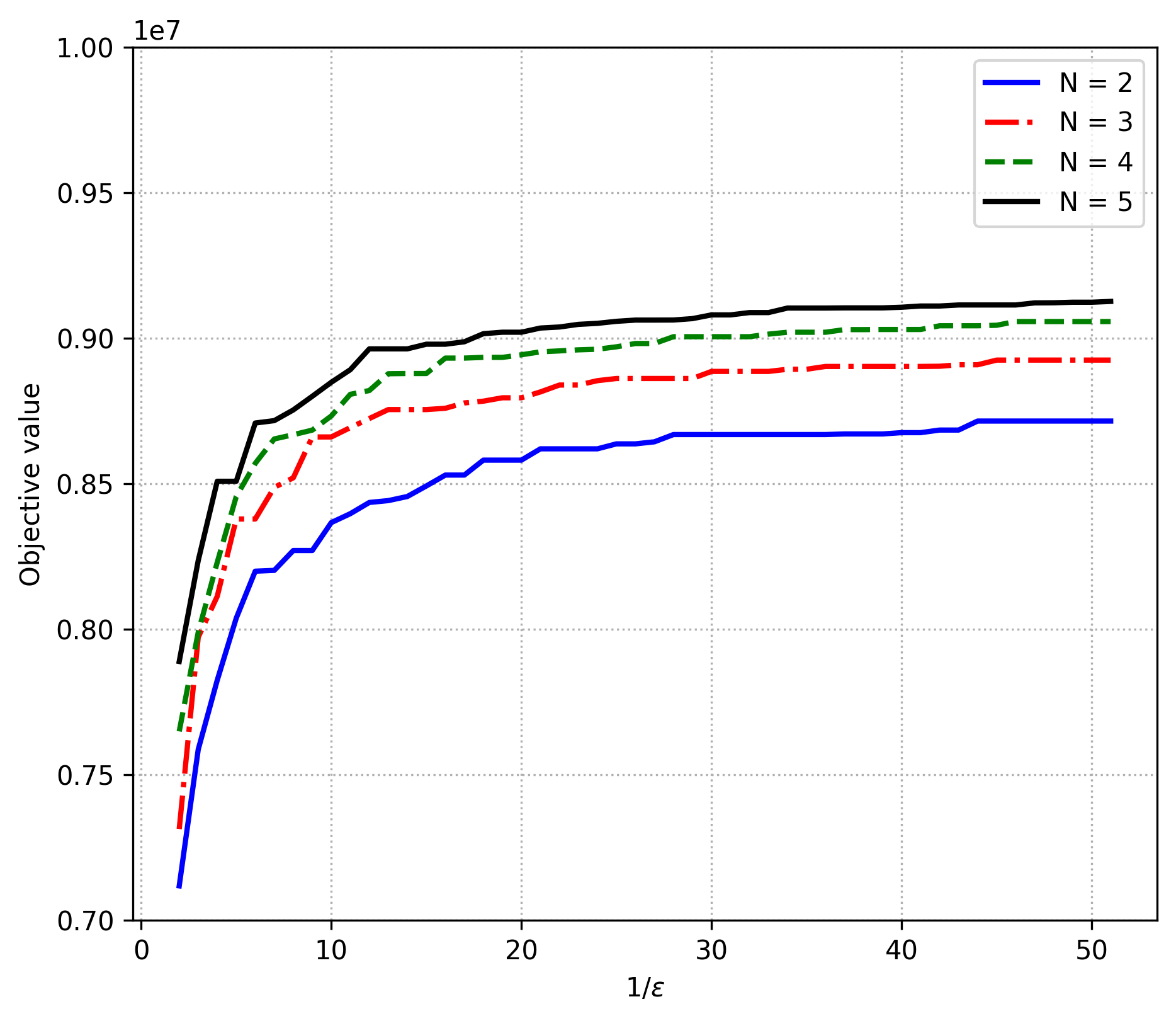}
  \caption{Objective value obtained by $\fptas$ versus $\epsilon$.}
  \label{fig:converge}
\end{figure}

First, we recall that $\epsilon$ is considered to be an accuracy parameter. As $\epsilon$ is small, the algorithm can obtain an accurate solution, but requires more execution time. To show how to properly select $\epsilon$, we change the value of $1/\epsilon$, and the objective values obtained by the $\fptas$ algorithm are shown in Fig.~\ref{fig:converge}. As seen, for a given number of users, the objective value increases with the increase of $1/\epsilon$, i.e., the decrease of $\epsilon$. Especially, the objective value seems to keep unchanged as $1/\epsilon>10$, i.e., $\epsilon<0.1$. As mentioned earlier, as $\epsilon$ is smaller, the execution time is longer. Therefore, for the trade-off between the objective value and the execution time, we select $\epsilon=0.1$ as a simulation parameter.  


\begin{figure}[h!]
  \centering
  \includegraphics[width=0.6\linewidth]{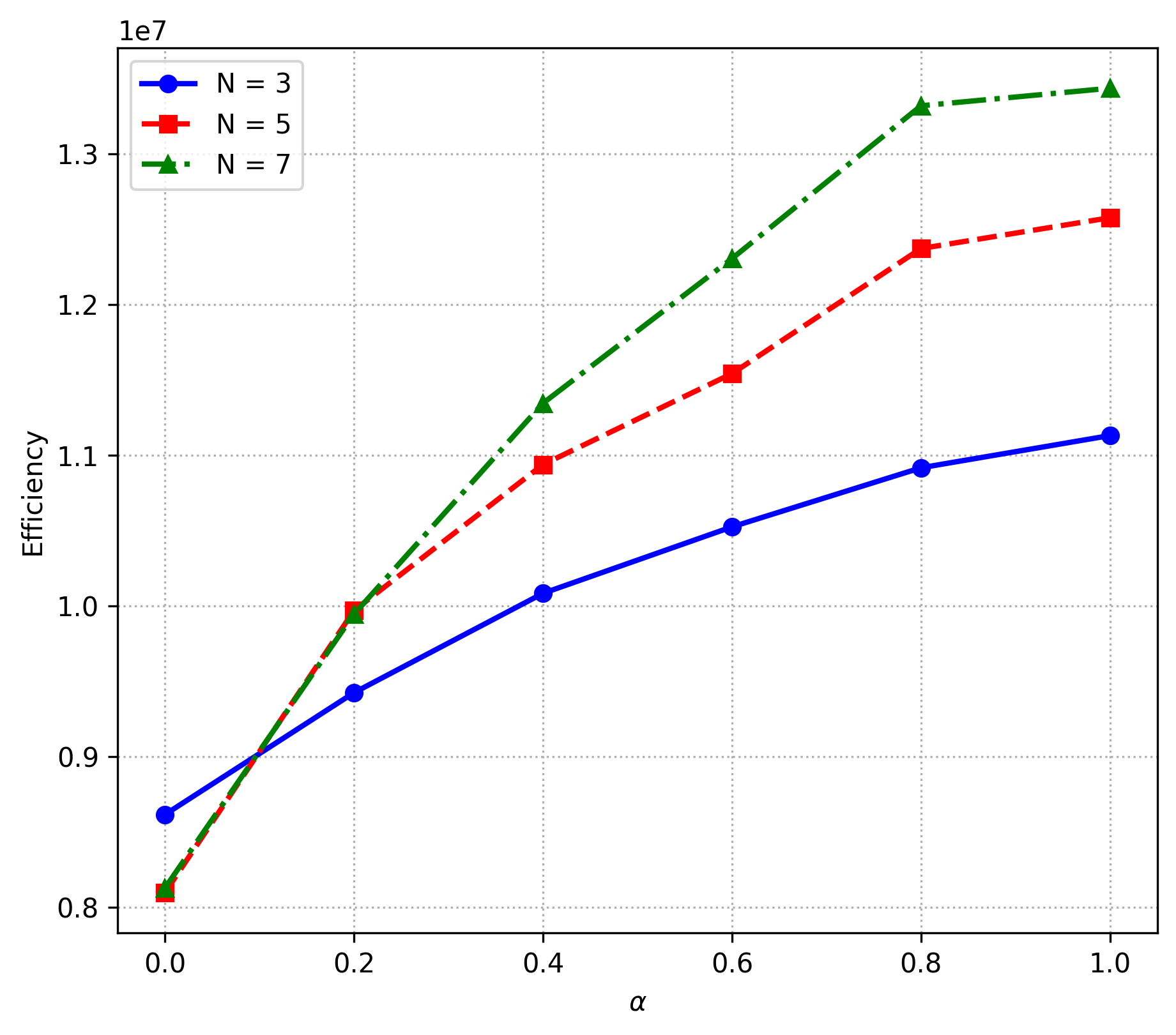}
  \caption{Efficiency versus weight parameter $\alpha$.}
  \label{fig:sub5}
\end{figure}

\begin{figure}[h!]
  \centering
  \includegraphics[width=0.6\linewidth]{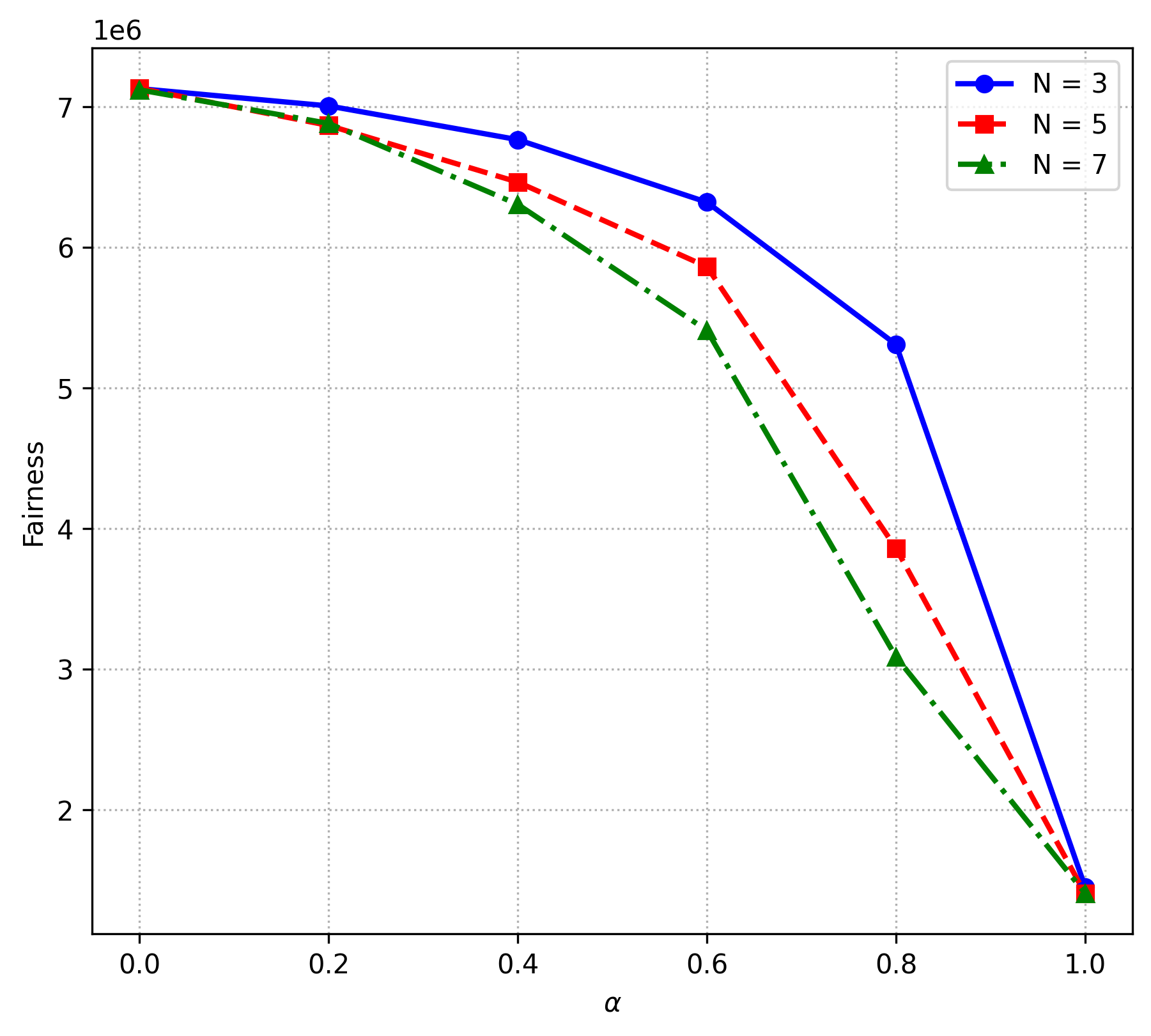}
  \caption{Fairness versus weight parameter $\alpha$.}
  \label{fig:sub4}
\end{figure}

\begin{figure}[h!]
  \centering
  \includegraphics[width=0.6\linewidth]{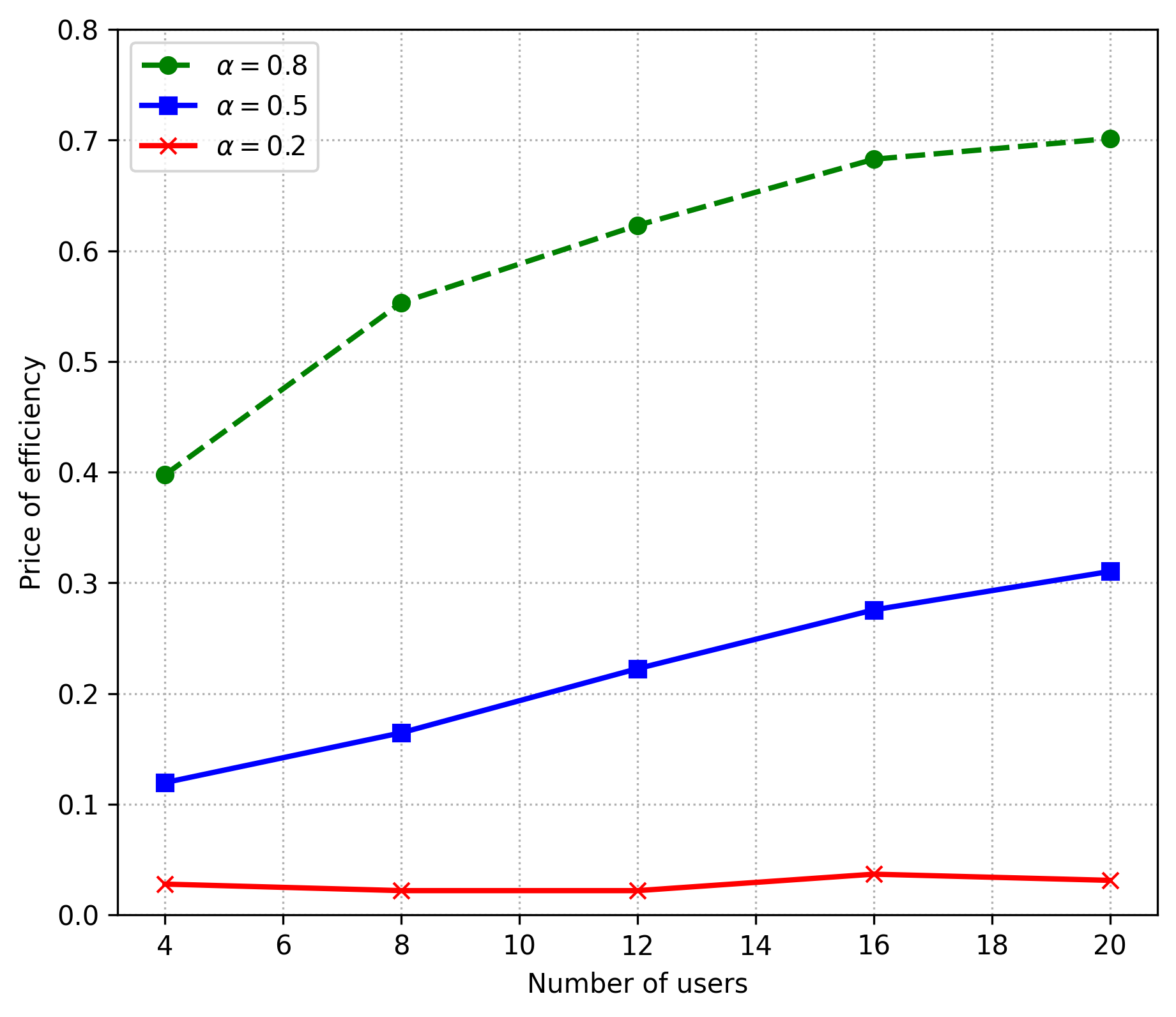}
  \caption{Price of efficiency of $\fptas$ versus the number of users and $\alpha$.}
  \label{fig:sub1}
\end{figure}

\begin{figure}[h!]
  \centering
  \includegraphics[width=0.6\linewidth]{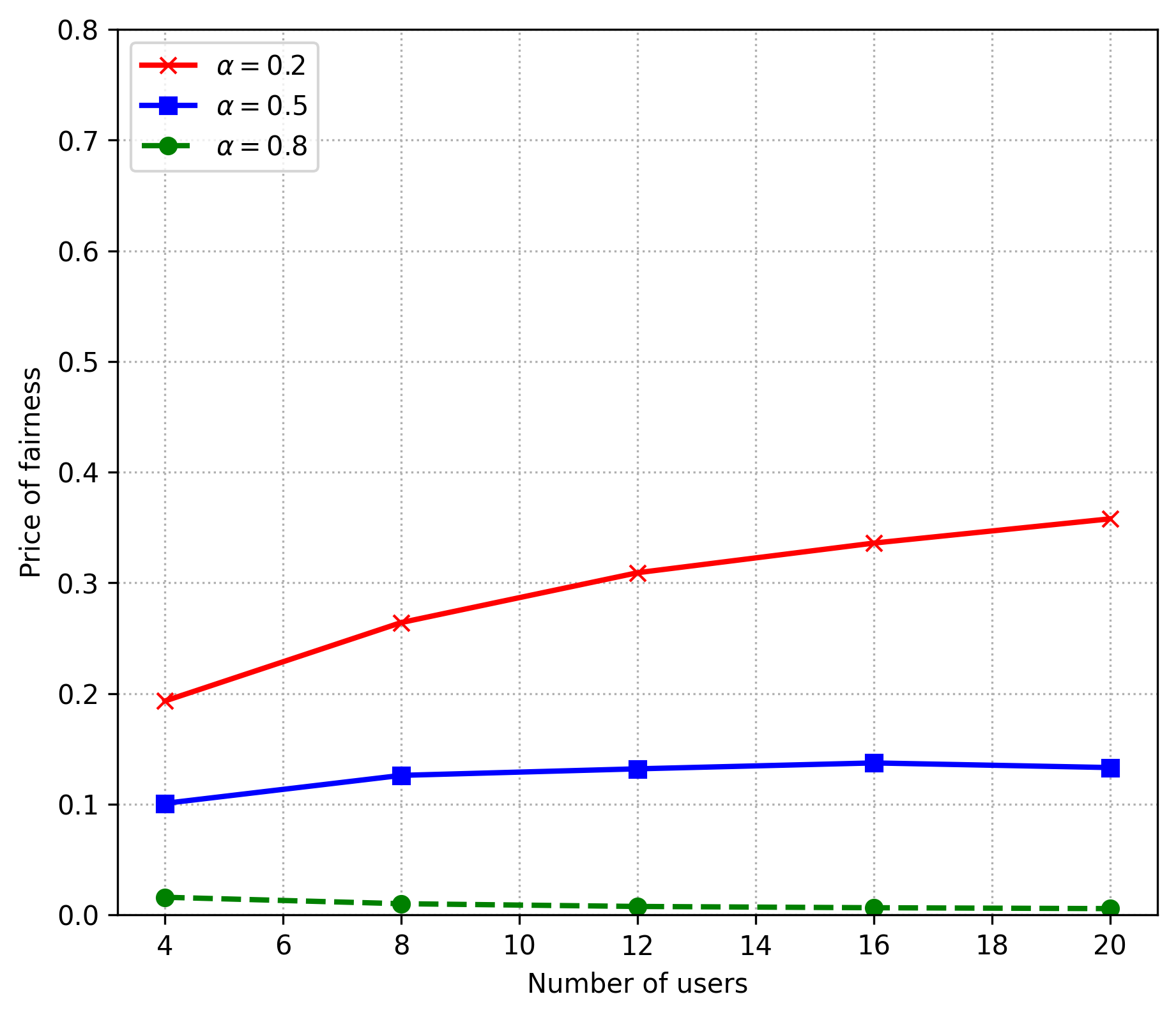}
  \caption{Price of fairness of $\fptas$ versus the number of users and $\alpha$.}
  \label{fig:sub2}
\end{figure}

\begin{figure}[h!]
  \centering
  \includegraphics[width=0.7\linewidth]{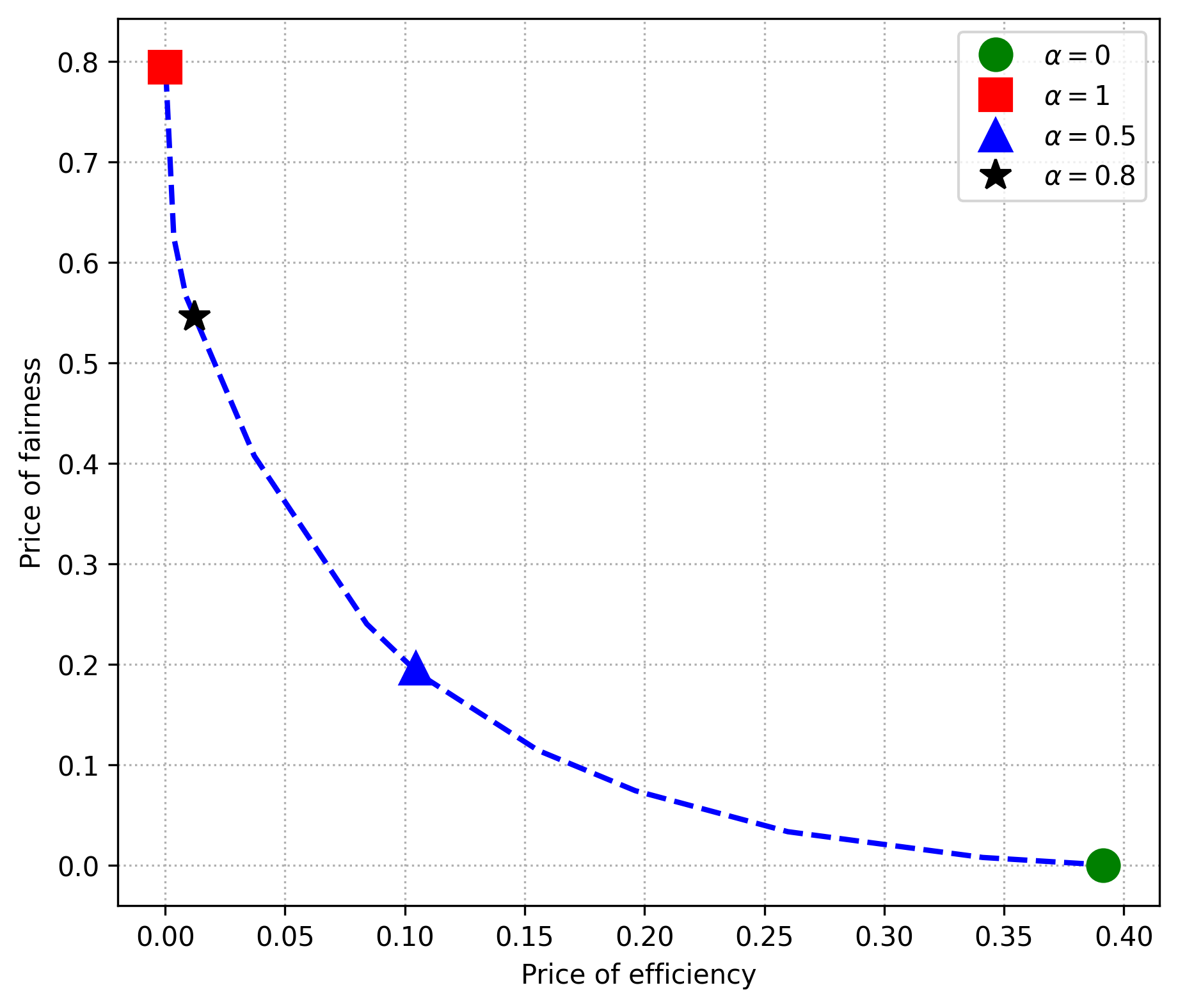}
  \caption{$\pof$ versus $\poe$ for $N=5$ users.}
  \label{fig:sub3}
\end{figure}

For the selection of the weight parameter, i.e., $\alpha$, this task is generally challenging due to the conflict between the efficiency and fairness.  Therefore, it is important and necessary to discuss the trade-off between the efficiency and fairness as $\alpha$ varies. The simulation results are shown in Figs.~\ref{fig:sub5},~\ref{fig:sub4},~\ref{fig:sub1},~\ref{fig:sub2}, and~\ref{fig:sub3}. As seen from Fig.~\ref{fig:sub5}, as $\alpha$ increases, the efficiency obtained by $\fptas$ increases. This is obvious since according to (\ref{FEO-obj}), as $\alpha$ increases, the $\fptas$ algorithm aims to improve the efficiency rather than the fairness. It can be further seen from Fig.~\ref{fig:sub5} that as $N$ increases, the efficiency increases. This is simply because of the property of logarithmic function used in the definition of both the sensing and communication functions. Based on (\ref{FEO-obj}), we can simply explain that the fairness obtained by $\fptas$ decreases as $\alpha$ increases as shown in Fig.~\ref{fig:sub4}. Furthermore, the fairness obtained by $\fptas$ decreases as the number of user increases. The reason is that given a fixed amount of bandwidth, the more users there are, the less utility the worst-off user has.

Now, we set $\alpha$ to some certain values and we discuss the the price of fairness and the price of efficiency, denoted by $\pof(\alpha)$ and $\poe(\alpha)$, respectively. In particular, we set $\alpha=0.2, 0.5, 0.8$. The simulation results are illustrated in Figs.~\ref{fig:sub1} and \ref{fig:sub2}. As seen from Fig.~\ref{fig:sub1}, as $\alpha$ increases, the $\poe$ increases. This is again consistent with the definition of the objective function, i.e., FEO. Also, $\poe$ increases with the number of users. However, given a small value of $\alpha$, i.e., $\alpha=0.2$, the $\poe$ is small, i.e., $\leq 0.05$, and seems to not change as $N$ varies. Given a value of $\alpha=0.2$, the weight of $0.8$ is associated with the fairness, which is quite close to $1$ at which we achieve the full fairness. Consequently, the change in the loss of fairness, i.e, $\poe$, is small and does not much depend on the number of users. However, as shown in Fig.~\ref{fig:sub2}, given $\alpha=0.2$, $\pof$ significantly increases as the number of users increases. In contrast to the case with $\alpha=0.2$, given $\alpha=0.8$, $\pof$ is small, i.e., $0.01$, and keeps unchanged over the number of users, but $\poe$ considerately increases as $N$ increases.


The results in Figs.~\ref{fig:sub1} and~\ref{fig:sub2} show that there is a significant loss in fairness or efficiency as $\alpha=0.2$ or $0.8$, respectively. The results further imply that $\alpha=0.5$ can be an appropriate selection to balance the trade-off between $\pof$ and $\poe$. To further demonstrate this point, Fig.~\ref{fig:sub3} shows $\pof$ versus $\poe$ in the case of $N=5$ users. We define that $\alpha=0$ is corresponding to the full fairness and $\alpha=1$ is corresponding to the full efficiency. As seen, if we choose $\alpha=0$, i.e., the full fairness, the loss of efficiency is nearly $80\%$ compared with full efficiency. Also, if we choose $\alpha=1$, i.e., the full efficiency, the loss of fairness is nearly $40\%$ compared with the full fairness. Meanwhile, if we select $\alpha=0.5$, there is only $10\%$ loss in the fairness and nearly $20\%$ loss in the efficiency compared with the full fairness and full efficiency, respectively. 



In summary, we set the accuracy parameter as $\epsilon=0.1$ and the weight parameter as $\alpha=0.5$ for the performance comparison, which is presented in the next section.
\subsection{Performance Comparison}
\label{sec:performance}

\begin{figure}[h!]
  \centering
  \includegraphics[width=0.6\linewidth]{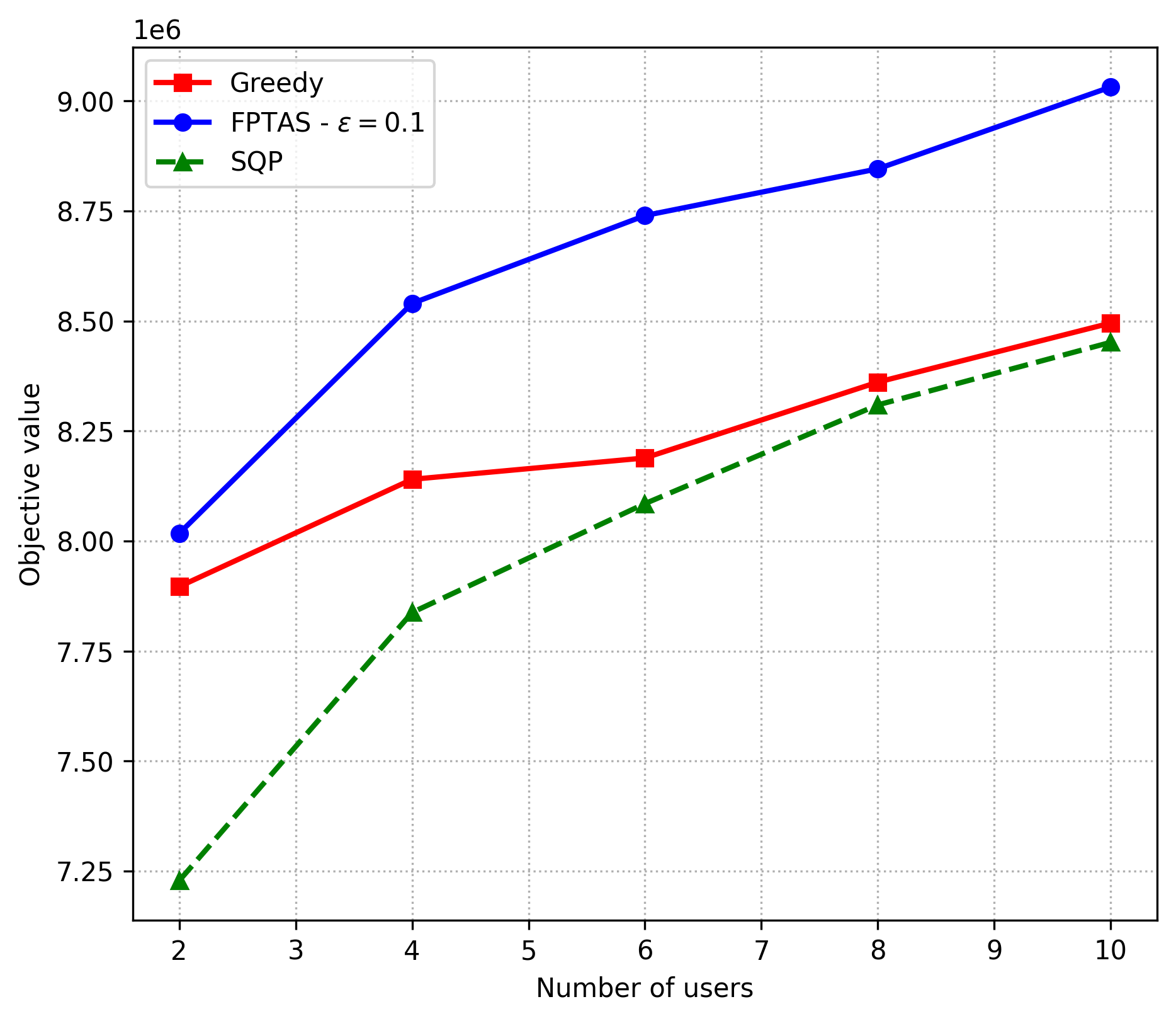}
  \caption{Performances obtained by $\fptas$, $\greedy$, and $\sqp$ algorithms versus the number of users.}
 \label{fig:obj-p2}
\end{figure}

\begin{figure}[h!]
  \centering
  \includegraphics[width=0.6\linewidth]{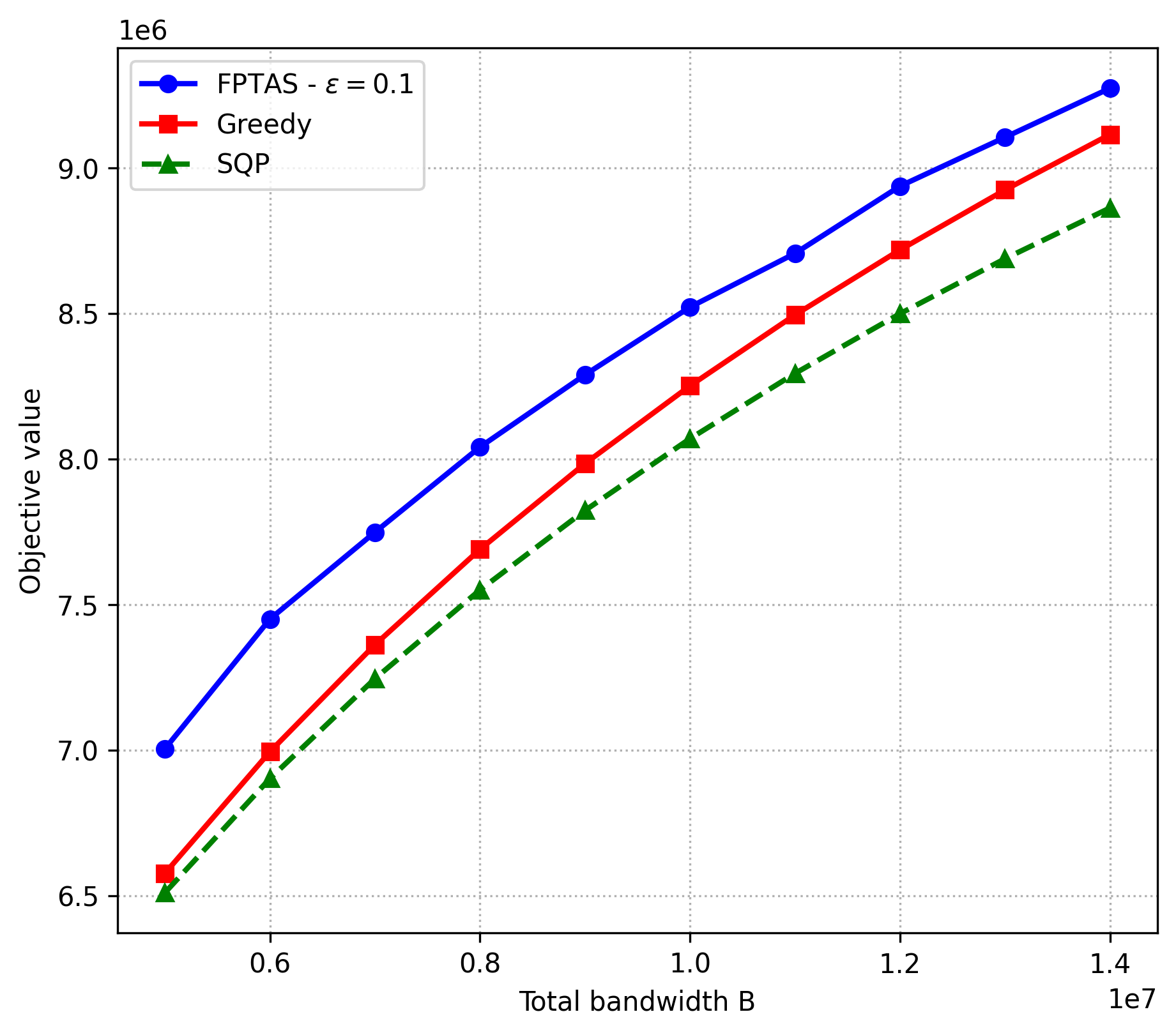}
  \caption{Objective value versus the total bandwidth.}
  \label{fig:change-B}
\end{figure}

\begin{figure}[h!]
  \centering
  \includegraphics[width=0.6\linewidth]{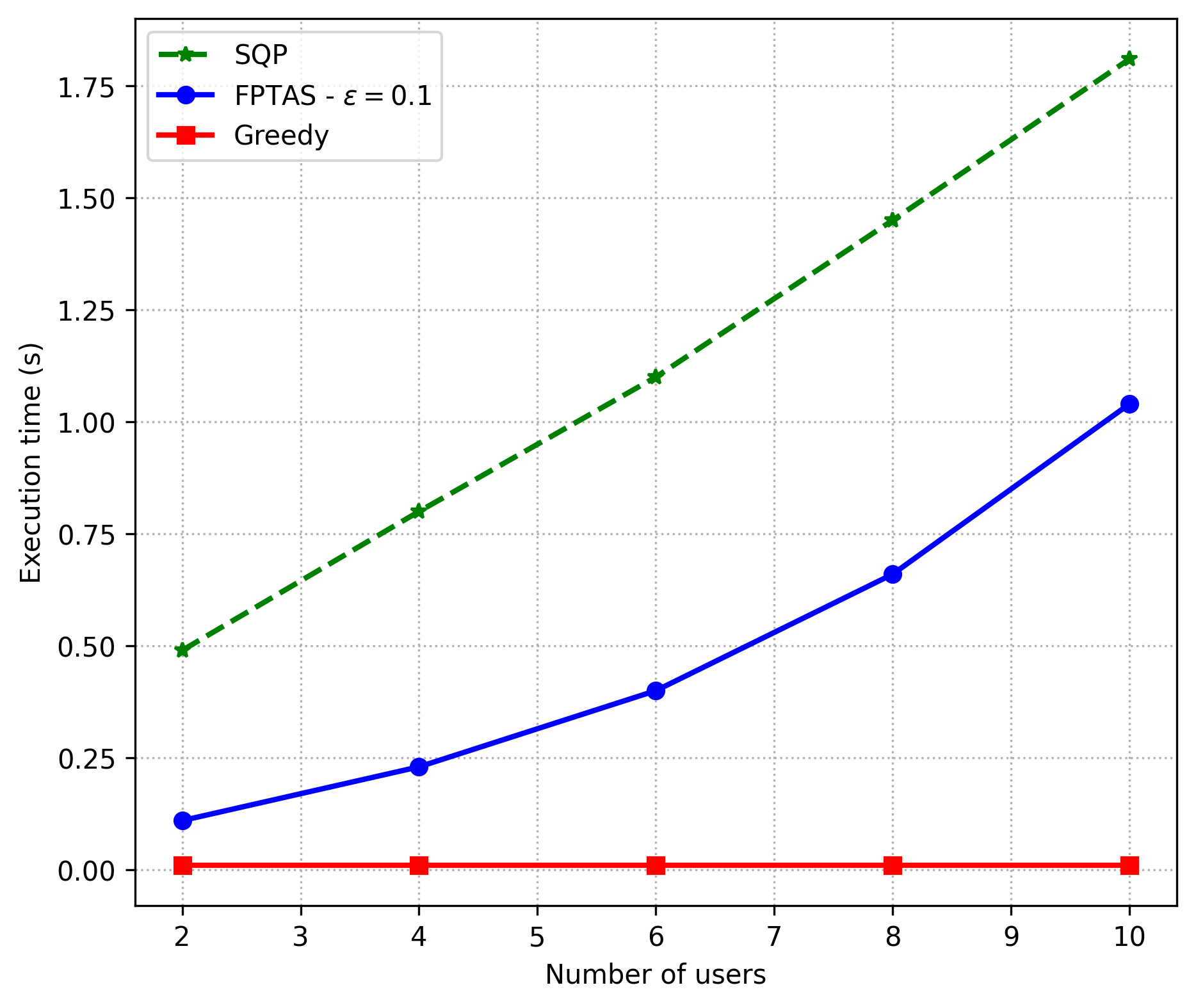}
  \caption{Execution time of $\fptas$, $\greedy$, and $\sqp$.}
 \label{fig:run-p2}
\end{figure}

Our work aims to maximize the system performance with the objective function including the efficiency and fairness. Moreover, the sensing function as a radar performs the target tracking, and thus the execution time of the algorithms needs to be considered. Therefore, in this section, we compare the algorithms in terms of the objective value and execution time, which are shown in~Figs.~\ref{fig:obj-p2}, \ref{fig:run-p2}, \ref{fig:change-B}, and Table~\ref{tab:compare}. Note that the objective value is the value of objective function $\cF(\bx)$ in (\ref{FEO-obj}), which is the weighted sum of the achievable efficiency and fairness. 

Fig. \ref{fig:obj-p2} illustrates the performances obtained by the $\fptas$, $\greedy$, and $\sqp$ algorithms when the number of users vary. As seen, the objective values obtained by the proposed algorithms, i.e., $\fptas$ and $\greedy$, are higher than those obtained by the $\sqp$ algorithm over the number of users. This means that the total efficiency and fairness obtained by the proposed algorithms are higher than those obtained by the baseline algorithm. 

Next, we examine how the performances obtained by the $\fptas$, $\greedy$, and $\sqp$ algorithms vary as the total bandwidth $B$ varies. Here, we fix the number of users to $N=5$ and vary the value of $B$ in the range $[5\times 10^6, 1.4\times 10^7]$. As shown in Fig.~\ref{fig:change-B}, over the values of $B$, the objective values obtained by the $\fptas$ and $\greedy$ algorithms are always much higher than that obtained by the $\sqp$ algorithm. Moreover, as the total bandwidth $B$ increases, the objective values obtained by all the algorithms increase. This is due to fact that the objective function (i.e., defined in $\cF(\bx)$) is (strictly) increasing on its domain.

Finally, we compare the execution time of the algorithms, and the results are shown in Fig.~\ref{fig:run-p2}. As seen, the execution time of the proposed algorithms, e.g., $\greedy$, are much shorter than those of the $\sqp$. This is due to the fact that the $\greedy$ algorithm has a linear complexity of $O(N)$. It can be seen from Fig.~\ref{fig:run-p2} that the execution time of most of algorithms increases as the number of users $N$ increases. In particular, the increase of execution time of $\fptas$ is polynomial in $N$. This is consistent with our complexity analysis in (\ref{eq:execution-fptas}) that the execution time  is quadratic in $N$. Meanwhile, $\greedy$ has a linear complexity of $O(N)$, and thus the increase of execution time of $\greedy$ seems to remain unchanged as $N$ increases, when compared to that of $\fptas$.

\renewcommand{\arraystretch}{1.5}
\begin{table}[h]
 \centering
\begin{tabular}{cccccccc}
\hline\hline
\multirow{2}{*}{$N$} &      \multicolumn{3}{c}{$\fptas$}   & & \multicolumn{3}{c}{$\greedy$}  \\ \cline{2-4} \cline{6-8}  
                       & obj. &  & time &  & obj. &  &time \\ \cline{2-2} \cline{4-4}    \cline{6-6} \cline{8-8}  
$2$           &   $11\%$       && $4.5$    &&  $9.2\%$       && $180$   \\ 
$4$           &   $9.0\%$       && $3.4$     &&  $3.9\%$       && $180$  \\ 
$6$           &  $8.0\%$        && $2.8$    &&  $1.3\%$       && $180$   \\ 
$8$           &   $6.5\%$       && $2.2$     &&  $0.6\%$       && $180$  \\ 
$10$         &   $6.9\%$       && $1.8$     &&  $0.5\%$       && $180$    \\ \hline
\end{tabular}
\caption{A summary of the objective values and execution time obtained by $\fptas$ and $\greedy$, compared with the $\sqp$ algorithm.}
\label{tab:compare}
\end{table}

To clearly demonstrate how our proposed algorithms improve the performance compared with $\sqp$, we use Table \ref{tab:compare}. In the table, the ``obj.'' column shows the percentage of objective value that $\fptas$ and $\greedy$ improve compared with $\sqp$, and the ``time'' column presents how many times the proposed algorithms executes faster than $\sqp$. As seen, the $\fptas$ algorithm can improve the objective value up to $11\%$ compared with $\sqp$, while keeping the execution time faster. Especially, the execution time of $\greedy$ is $180$ times faster compared with $\sqp$ while improving the objective value up to $9.2\%$.
 
 The results demonstrate that compared with the baseline algorithm, our proposed algorithms better adapt the low delay requirements of the radar systems. Clearly, our proposed algorithms are scalable. Furthermore, the $\fptas$ algorithm can be a good choice if we aim to achieve a higher objective value within reasonable execution time, i.e., up to one second for $N=10$ users. Meanwhile, $\greedy$ will be a better alternative if we need short execution time, i.e., less than $1\%$ second.


\section{Conclusion}
\label{sec:conclusion}
In this paper, we have investigated the bandwidth allocation problem in the DJSC system. First, we have formulated the optimization problem that aims to optimize bandwidth allocation to the sensing and communication functions of the JSC users. The objective is to maximize the total sensing performance, i.e., estimation rate, communication performance, i.e., communication data rate, and the max-min fairness of all the users. To solve the non-convex problem, we propose the polynomial time approximation algorithm that is able to find a near-optimal solution. To further reduce the execution time, we propose to use the heuristic algorithm that performs the bandwidth allocation to the JSC users in a greedy manner. We have provided simulation results to demonstrate the improvement and effectiveness of the proposed algorithms, compared with the active-set sequential quadratic programming algorithm, which is known as the currently best algorithm for solving non-linear optimization. The simulation results further showed that the heuristic algorithm can be a more suitable solution to the DJSC systems when the short execution time is required.

\section*{Appendix}

\subsection{The non-concavity of $\cF_p$}
\label{sub-obs:1}

We prove that $\cF_p$ is neither concave nor convex for $p\ge 2$. Let $\Gamma_i(x)=(f_i(x)+g_i(x))^p$ over the interval $[0,B]$, where $p\in \mathbb{R}\setminus \{0\}$, and the second order derivative of $\Gamma_i(x)$ can be determined as follows:
\begin{multline*}
\Gamma''_i(x)=-\frac{p}{\ln 2}\left(\dfrac{2\nu_i^2}{T_{\text{pri}}\left(1+\nu_ix\right)^2}+\dfrac{\tau_i^2}{x\left(x+\tau_i\right)^2}\right)\times \\
\times\left(\frac{2}{T_{\text{pri}}}\ln\left(\nu_ix+1\right)+x\ln\left(\dfrac{\tau_i}{x}+1\right)\right)^{p-1}\\
+\frac{(p-1)p}{\ln 2}\left(\frac{2}{T_{\text{pri}}}\dfrac{\nu_i}{\nu_i x+1}-\dfrac{\tau_i}{\tau_i+x}+\ln\left(\dfrac{\tau_i}{x}+1\right)\right)^2\times\\
\times \left(\frac{2}{T_{\text{pri}}}\ln\left(\nu_i x+1\right)+x\ln\left(\dfrac{\tau_i}{x}+1\right)\right)^{p-2}.
\end{multline*}
For $p>1$, we can verify that $\Gamma''_i(x)$ may take both negative and positive values over $[0,B]$, and thus $\Gamma_i(x)$ is neither convex nor concave. Indeed, we consider a simple form of the function when $p=2$, $\tau_i=\nu_i=1$ and $T_{\text{pri}}=2$. Then, we can compute $\Gamma''_i(x)$ as follows:
\[
-\dfrac{2\left(\ln\left(x+1\right)-\ln^2\left(\frac{1}{x}+1\right)x^2+\left(\ln\left(\frac{1}{x}+1\right)-\ln^2\left(\frac{1}{x}+1\right)\right)x\right)}{x\left(x+1\right)\ln 2},
\] 
for which $\Gamma''_i(1/3)\cdot \Gamma''_i(1)<0$ holds.

\subsection{Trade-off between Efficiency and Fairness}
\label{trade-off}
\begin{figure}[h!]
  \centering
  \includegraphics[width=0.7\linewidth]{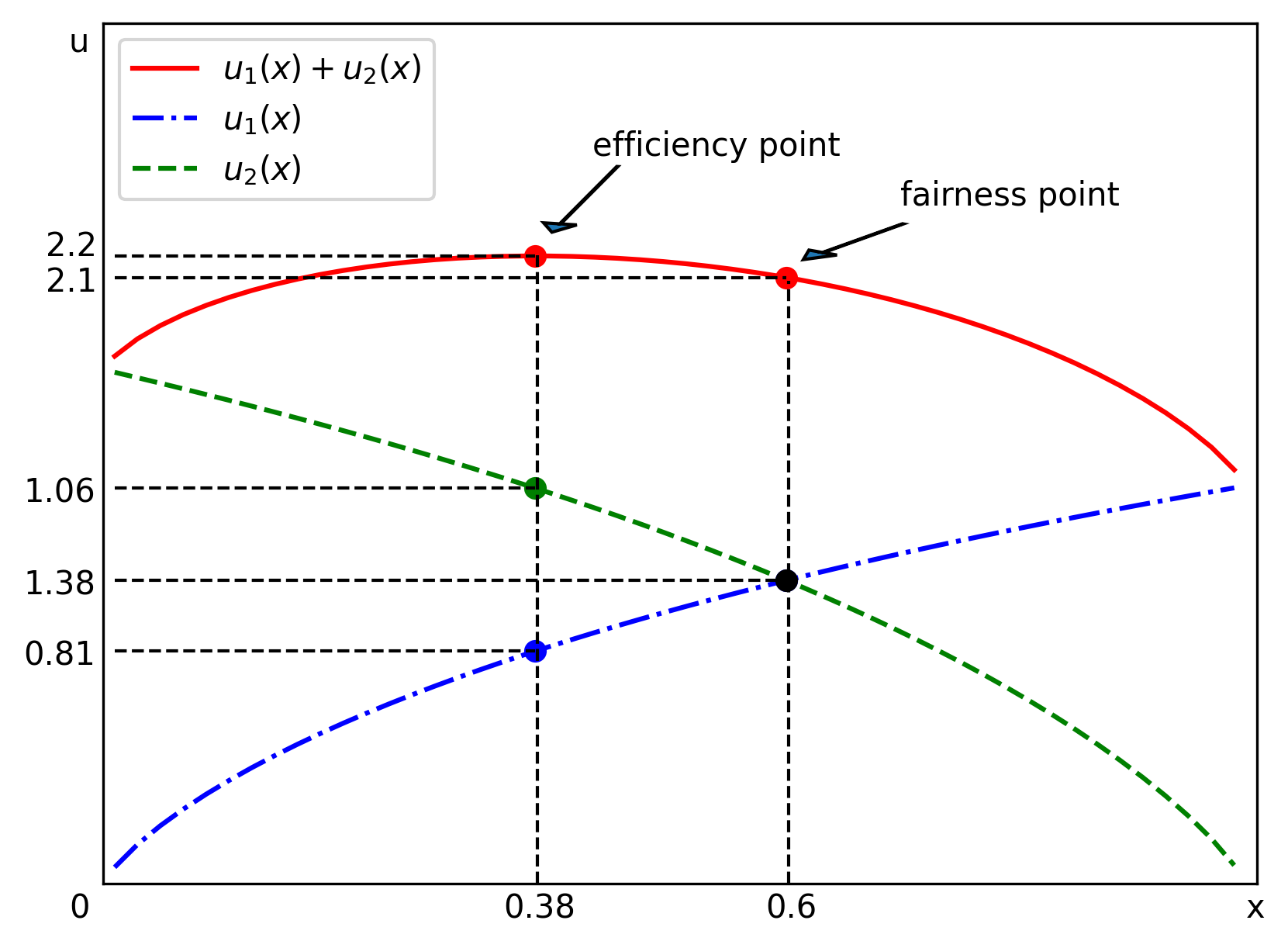}
  \caption{Trade-off between efficiency and fairness.}
  \label{fig:trade-off}
\end{figure}

Consider an example with two users whose utility functions are given as follows. W.l.o.g we can assume that the total bandwidth is $1$. Let 
\begin{align*}
u_1(x)=&~x\log_2(1+\frac{1}{x})+  \log_2(1+x),~\text{and}\\
u_2(x)=&~(1-x)\log_2(1+\frac{2}{1-x}) +\log_2(2-x),
\end{align*}
where $x\in(0,1)$ denotes the amount of bandwidth allocated to user $1$. Consider the case with $p=1$. Then, the maximum efficiency of the instance problem is computed as  $\max_{x\in(0,1)} \{u_1(x)+u_2(x)\}$. This is convex problem as the function is concave, and thus one can easily find its optimal point $x=0.38$, with the corresponding efficiency of $2.2$, as shown in  Fig~\ref{fig:trade-off}. Note that, at this point, the utilities of the two users are respectively $0.81$ and $1.06$, resulting in the fairness of $0.81$. On the other hand, it is seen that the maximum fairness is achieved at the intersection point of the two functions $u_1$ and $u_2$, which gives each user an equal utility of $1.38>0.81$. This follows the fact that we can improve efficiency only at the cost of increasing fairness and vice versa, making a trade-off between the efficiency and the fairness of bandwidth allocation.

\subsection{Proof of Lemma~\ref{claim:epsilon} }
Recall that a continuous function $u:\mathbb{R}\to\mathbb{R}$ is $c$-Lipschitz  if for all $x$ and $x'$ (in the domain of $u$), $|u(x)-u(x')|\le c|x-x'|$, and a sufficient condition for this is that $|u'(x)|\le c$ for all $x$. Note that $u_i'(0)=\infty$, and thus we need to restrict $u_i$ to an interval that is sufficiently away from $0$.

\begin{claim}
\label{claim:1}
We can choose $y_0>0$ such that the root $y_i^1$ (at which $u_i(y_i^1)=\frac{\epsilon L}{\sqrt[p]{N}}$) lies in the interval $[y_0,\overline\xi_i]$.
\end{claim}
\begin{proof}
 Let $c_i>1$ be a constant and $y_1:=\frac{\epsilon L}{2c_i\sqrt[p]{N}}$. Then, based on the definition of $u_i$, we can express $u_i(y_1)$ as follows:
\begin{multline*}
u_i(y_1)= f_i(y_1)+g_i(y_1)= \frac{1}{2T_{\text{pri}}}\cdot \log_2\left(1+\frac{\nu_i \epsilon L}{2c_i \sqrt[p]{N}}\right)\\
+\frac{\epsilon L}{2c_i \sqrt[p]{N}}\log_2\left(1+\dfrac{c_i\sqrt[p]{N}\tau_i}{\epsilon L}\right).
\end{multline*}
If $\dfrac{\sqrt[p]{N}\tau_i}{\epsilon L}\le 2$, then $c_i=3$ and $g_i(y_0)\le\frac{\epsilon L}{\sqrt[p]{N}}\cdot\frac{\log_2 6}{3}<\frac{\epsilon L}{\sqrt[p]{N}}$. Otherwise, let $c_i=2\log_2 \frac{\sqrt[p]{N}\tau_i}{\epsilon L}$. Then
\begin{align*}
g_i(y_1)&\le\frac{\epsilon L}{2c_i\sqrt[p]{N}}\log_2\left((1+c_i)\dfrac{\sqrt[p]{N}\tau_i}{\epsilon L}\right)\\&=\frac{\epsilon L}{2\sqrt[p]{N}}\left(\frac{\log_2(1+c_i)}{c_i}+\frac{1}{e}\right)<\frac{\epsilon L}{2\sqrt[p]{N}},
\end{align*}
using the inequality $\log(1+ea)<(e-1)a$, which is valid for all $a\ge1$. 

Otherwise, given a positive value of $y_2$ with $y_2<\frac{1}{\tau_i}(2^{\frac{T_{\text{pri}}\epsilon L}{\sqrt[p]{N}}}-1)$, we have $f_i(y_2)<\frac{\epsilon L}{2\sqrt[p]{N}}$. Due to the monotone increasing of $f_i(x)+g_i(x)$, it is enough to choose $0<y_0<\min\{y_1,y_2\}$ such that $f_i(y_0)+g_i(y_0)<\frac{\epsilon L}{\sqrt[p]{N}}$. Hence, we have $u_i(y_0)< \frac{\epsilon L}{\sqrt[p]{N}}$ and the root $y_i^1$ (at which $u_i(y_i^1)=\frac{\epsilon L}{\sqrt[p]{N}}$) lies in the interval $[y_0,\overline\xi_i]$.~\end{proof}

 Now, for any $x$ in the interval $[y_0,\overline\xi_i]$, we have
\begin{align*}
u_i'(x)&=f_i'(x)+g'_i(x)\\
&~= \frac{\nu_i}{2T_{\text{pri}}} \cdot \dfrac{1}{1+\nu_i x}+ \log_2\left(\dfrac{\tau_i+x}{x}\right)-\dfrac{\tau_i}{\tau_i+x}\\
&~\le    \frac{\nu_i}{2T_{\text{pri}}}+ \log_2\left(\dfrac{\tau_ic_i\sqrt[p]{N}}{\epsilon L}+1\right)\\
&~\le  \frac{\nu_i}{2T_{\text{pri}}}+\dfrac{\tau_ic_i\sqrt[p]{N}}{\epsilon L}.
\end{align*}
By the Lipschitz condition mentioned above, it follows that we should select 
\[
\epsilon_i:= \frac{\epsilon^2L}{\sqrt[p]{N}}: \left( \frac{\nu_i}{2T_{\text{pri}}}+\dfrac{\tau_ic_i\sqrt[p]{N}}{\epsilon L}\right),
\]
such that
\begin{align*}
|u_i(\tilde x_i^j)-u_i(x_i^j)|\le \left( \frac{\nu_i}{2T_{\text{pri}}}+\dfrac{\tau_ic_i\sqrt[p]{N}}{\epsilon L}\right)|\tilde x_i^j-x_i^j|\le \epsilon u_i(x_i^j).
\end{align*}

\subsection{Proof of Lemma~\ref{cl:discrete} }
\begin{proof}
Let $\epsilon_0=\min_{i\in\cN}\epsilon_i$. The discretization asks to find, for each $i\in\cN$, a solution $\tilde x_i^j$ to the equation $u_i(x)=u_i^j$, for $j=0,1,\ldots,K_i$. Hence, the overall complexity of this process, $\tim_{\text{discrete}}$, is bounded by $O(T)$, where 
\begin{align*}
T=&~{\sum}_{i\in\cN}  \left( \log_2 \frac{u_i(\overline\xi_i)}{\epsilon_i} \cdot \log_{(1+\epsilon)}\frac{\sqrt[p]{N} u_i(\overline\xi_i)}{\epsilon L}\right)\\
\le&~ \frac{1}{\log_2(1+\epsilon)}\cdot {\sum}_{i\in\cN} \left(\log_2 \frac{u_i(\overline\xi_i)}{\epsilon_i} \cdot \log_2\frac{u_i(\overline\xi_i)}{\epsilon}\right)\\
\le &~ \frac{1}{\epsilon}\cdot \log_2 \frac{1}{\epsilon}\cdot \log_2 \frac{1}{\epsilon_0}\cdot N\cdot {\max}_{i\in\cN} \left\{\log^2_2 u_i(\overline\xi_i)\right\},
\end{align*}
where the first inequality follows from the fact that $L= (\sum_{i\in\cN} (u_i(\underline\xi_i))^p)^{1/p}\ge \sqrt[p]{N}$ and $u_i(\underline\xi_i)\ge 1$ for all $i$. while the second inequality is because $\epsilon_0\le \epsilon_i$ for all $i$. On the other hand, from the definition of $\epsilon_i$ we have that
\[
\log_2 \left(\frac{1}{\epsilon_i}\right)\le  \log_2 \left(\frac{\frac{\nu_i}{2T_{\text{pri}}}+\tau_ic_i}{\epsilon^3}\right)\le \log_2 \left(\frac{\nu_i}{2T_{\text{pri}}}+\tau_ic_i\right)\cdot \log_2 \frac{1}{\epsilon^3}.
\]
Therefore, it holds that 
\[
\log_2 \left(\frac{1}{\epsilon_0}\right)\le \log_2 \frac{1}{\epsilon^3}\cdot  \max_{i\in\cN} \left\{\log_2 \left(\frac{\nu_i}{2T_{\text{pri}}}+\tau_ic_i\right)\right\}.
\]
This follows that
\begin{align*}
T\le&~ \frac{3}{\epsilon}\log^2_2 \frac{1}{\epsilon}\cdot N\cdot \omega,
\end{align*}
where $\omega =   {\max}_{i\in\cN} \left\{\log^2_2 u_i(\overline\xi_i)\cdot \log_2 \left( \frac{\nu_i}{2T_{\text{pri}}}+\tau_ic_i   \right)\right\}$. Also, note that $c_i=\max\{2,2\log_2 \frac{\sqrt[p]{N}\tau_i}{\epsilon L}\}\le \max\{2,2\log_2 \frac{\tau_i}{\epsilon}\}$, as $\sqrt[p]{N}\le L$. This completes the proof of Lemma~\ref{cl:discrete}.
~\end{proof}



\subsection{Proof of Lemma~\ref{lem:complexity-DP} }

\begin{proof}
It is not difficult to see that the execution time of {\bf DP} is $\tim_{\textsc{DP}}=O(N'\cdot \sum_{i\in\cN}|D_i|)=O(\frac{1}{\epsilon}N\cdot \sum_{i\in\cN} K_i)$, where $K_i\le 1+\log_{(1+\epsilon)}\frac{\sqrt[p]{N} u_i(\overline\xi_i)}{\epsilon L}\le 1+ \frac{1}{\epsilon}\log_2\frac{1}{\epsilon}\log_2 u_i(\overline\xi_i)$. Hence, $\tim_{\textsc{DP}}=O(\frac{1}{\epsilon^2}\log_2\frac{1}{\epsilon}N^2\max_{i\in\cN}\{\log_2 u_i(\overline\xi_i)\})$.~\end{proof} 

\bibliographystyle{IEEEtran}
\bibliography{IRS_database}{}

\end{document}